\documentclass[iop, revtex4]{emulateapj} 
\usepackage{apjfonts}
\usepackage{morefloats}
\usepackage{natbib}
\usepackage{epstopdf}
\usepackage{graphicx}
\usepackage[pdfa]{hyperref}
\usepackage{filecontents}
\usepackage{float}
\usepackage{color}
\shorttitle{Rigorous lens model of Abell 2744}
\begin{document}

\bibliographystyle{apj}

\title{A Rigorous Free-Form Lens Model of Abell 2744 to Meet the Hubble Frontier Fields Challenge}

\author{Daniel Lam\altaffilmark{1},
Tom Broadhurst\altaffilmark{2,3},
Jose M. Diego\altaffilmark{4},
Jeremy Lim\altaffilmark{1},
Dan Coe\altaffilmark{5},
Holland C. Ford\altaffilmark{6},
Wei Zheng\altaffilmark{6}
}

\altaffiltext{1}{Department of Physics, The University of Hong Kong, Pokfulam Road, Hong Kong}
\altaffiltext{2}{Department of Theoretical Physics, University of Basque Country UPV/EHU, Bilbao, Spain}
\altaffiltext{3}{IKERBASQUE, Basque Foundation for Science, Bilbao, Spain}
\altaffiltext{4}{Instituto de F\'isica de Cantabria, CSIC-Universidad de Cantabria, 39005 Santander, Spain}
\altaffiltext{5}{Space Telescope Science Institute, Baltimore, MD, USA}
\altaffiltext{6}{Dept. of Physics and Astronomy, Johns Hopkins university, Baltimore, Maryland, USA}

\setcounter{footnote}{11}

\begin{abstract}

Hubble Frontier Fields (HFF) imaging of the most powerful lensing clusters provides access to the most magnified distant galaxies. 
The challenge is to construct lens models capable of describing these complex massive, merging clusters so that individual lensed systems can be reliably identified and their intrinsic properties accurately derived. 
We apply the free-form lensing method (WSLAP+) to A2744, providing a model independent map of the cluster mass, magnification, and geometric distance estimates to multiply-lensed sources. 
We solve simultaneously for a smooth cluster component on a pixel grid, together with local deflections by the cluster member galaxies. 
Combining model prediction with photometric redshift measurements, we correct and complete several systems recently claimed, and identify 4 new systems - totalling 65 images of 21 systems spanning a redshift range of 1.4$<z<$9.8. 
The reconstructed mass shows small enhancements in the directions where significant amounts of hot plasma can be seen in X-ray. 
We compare photometric redshifts with "geometric redshifts", finding a high level of self-consistency. 
We find excellent agreement between predicted and observed fluxes - with a best-fit slope of 0.999$\pm$0.013 and an RMS of $\sim0.25$ mag, demonstrating that our magnification correction of the lensed background galaxies is very reliable. 
Intriguingly, few multiply-lensed galaxies are detected beyond $z\simeq 7.0$, despite the high magnification and the limiting redshift of $z\simeq 11.5$ permitted by the HFF filters. 
With the additional HFF clusters we can better examine the plausibility of any pronounced high-z deficit, with potentially important implications for the reionization epoch and the nature of dark matter.

\end{abstract}


\clearpage

\section{Introduction}

Cluster lensing has two major advantages over field surveys for exploring galaxy formation. 
In addition to the magnification boost, it is actually possible to derive source distances purely geometrically from the relative angles between sets of counter-images. 
This ability provides a very welcomed check of redshifts derived photometrically, particularly at high redshift where detection is weakest and restricted to fewer passbands.
For this purpose an accurate lens model is required, based on many sets of multiply-lensed images and ideally sampling a wide range of source distances. 
The angles through which light is deflected scale with increasing source distance behind a given lens. 
Distances derived this way can then be converted via cosmological parameters to source redshifts and compared with independently derived photometric redshifts. 
This method has been established using the large lensing cluster A1689, where geometric distances provided a self-consistency check of the lens model \citep{broadhurst05, limousin07, diego14}.

Currently, efficient detection of high redshift galaxies is best achieved using the IR channel of the HST Wide-Field Camera-3 (WFC3), supported by the Spitzer Space Telescope's Infrared Array Camera (IRAC) in the mid-IR. 
This combination is now generating statistically useful samples of dropout galaxies to $z\simeq 8$ in several independent deep field surveys \citep{oesch10, oesch13, ellis13, finkelstein13, schmidt14}. 
The highest redshift galaxies reliably claimed at present has been discovered in the recently completed CLASH Hubble Treasury program \citep{postman12} despite minimal exposures of only one orbit in the NIR bands. 
Although the principal aim of the CLASH program is to establish a representative distribution of equilibrium mass profiles for relaxed clusters, several high magnification clusters have been included in the hope of striking relatively bright examples of distant magnified galaxies. 
The highest redshift claimed is a triply-lensed small round object that lies an estimated redshift of z$\simeq$10.7 \citep{coe13}, followed by a similar object at $z=9.6\pm0.2$ \citep{zheng12}. 
Both these objects lie behind high magnification clusters defined by the CLASH program.

These discoveries have led to a dedication of a large Hubble ``Frontier Field'' (HFF) program of deep optical-NIR imaging to utilise cluster lensing for the purpose of discovering statistical samples of even higher redshift galaxies (Hubble Deep Fields Initiative 2012 Science Working Group Report\footnote{http://www.stsci.edu/hst/campaigns/frontier-fields/HDFI\_SWGReport2012.pdf}). 
In fact, since the phenomenon of cluster lensing was first appreciated, it has consistently provided record breaking redshifts and relatively bright images useful for spectroscopy \citep{soucail87, kneib93, frye98, bradley08, coe13}. 
The targets chosen for the HFF include some of the most magnifying clusters known, caught in the process of merging and therefore having complex mass distribution. 
These systems are characterised by large Einstein radii and shallow mass distributions, established in earlier Hubble imaging, principally from the ACS/GTO program \citep{ford03} and by the recently completed CLASH survey \citep{postman12}. 
The largest known of these lenses is MACS J0717.5+3745(z=0.55), which has an elongated critical area equivalent to an Einstein radius of 55\arcsec (at z=2.5) as originally uncovered by \citet{zitrin09a}.
This cluster was established by the CLASH survey to be the most massive known cluster at $z>0.5$ when weak lensing is added to derive the full virial mass distribution from lensing \citep{medezinski13}. 
Similarly, the merging cluster MACS J1149.5+2223(z=0.54) has a very shallow unrelaxed mass profile (Zintrin \& Broadhurst 2009) from adjacent overlapping substructures, forming a large critically lensed central area with very high magnification. 
For MACS J0416.1- 2403, long complex critical curves have been uncovered in the CLASH survey \citep{zitrin13} with multiple linear substructures.  
More recently we visited this cluster but using HFF data confirming the findings of \citet{zitrin13} \citep{diego14b}.
In the case of A2744, which is the subject of the work presented here, the impressive scale of lensing and the dynamical interaction of several mass components were reported by \citet{merten11}, offsets can be seen between the lensing mass components and the complex X-ray morphology, indicating collisions are on-going between perhaps three cluster sized objects. 
Out of the six clusters selected for the new Hubble Frontier Fields program, four have already been imaged to shallower depths by Hubble as part of the CLASH survey. 
Several sets of multiply lensed images have been catalogued for each cluster, having reliable photometric redshifts based on 16 overlapping optical/NIR filters and additional Spitzer data in the Mid-IR. 

The HFF program is in fact the first deep campaign with Hubble to utilise cluster lensing for studying galaxy formation. 
The high magnification clusters selected combined with long exposures in the near-infrared (NIR) is anticipated to extend the detection of galaxies to $z < 13$, corresponding to the effective limiting dropout redshift of the NIR filters employed by the HFF. 
The magnified flux limit for small sources is substantially fainter than the comparable long integrations of the deep field surveys. 
Supporting X-ray and Mid-IR data will be provided by additional deep Chandra and Spitzer satellite imaging, thereby extending the scientific interest in this uniquely deep dataset. 
In addition a good case for very deep complementary radio imaging with the extended-VLA (EVLA) can be made for these targets, given the potential serendipity value in the relatively unexplored radio regime at high redshift.

The improving quality of lensing data for clusters imaged with the HST encourages improved lens modelling to take advantage of the increasing constraints on the distribution of dark matter. 
Several lens models are already made publicly available, and can be catagorised as either parametric or free-form. 
Parametric modelling, which uses light as a rough guide for where to place masses, is best suited to virialized clusters \citep{halkola06, limousin07}, and may be extended to accommodate obvious bimodal substructure. 
In general, however, the complexities of massive merging clusters such as those chosen for the HFF program will require the definition of several new parameters for each additional model halo, with the choice of location being less than objective. 
The extent to which such modelling can capture the real complexities of tidally distorted dark matter during extreme gravitational encounters has prompted us to look harder at the possibility of grid solutions, whereby the lens plane can be represented on either a uniform or an adaptive grid. 
In early non-parametric studies, the uniform grid lens models were not accurate enough for identifying new sets of multiple images because they did not have high enough resolution to capture the local perturbing effects of cluster galaxies. 
A huge improvement has recently been achieved in this approach by incorporating the observed member galaxies along with the smoother and more distributed mass in a uniform grid. 
This then allows meaningful solutions to be found as the small scale deflections and additional multiple images locally generated by the member galaxies can be accounted for. 
This approach generates lens models that are sufficiently accurate and self-consistent for the identification of multiple systems, so that physically plausible mass distributions can be derived free of model assumptions. 
This has been demonstrated with both simulated data \citep{sendra14} and actual observation on a relaxed cluster (Abell 1689, \citet{diego14}). 
Although only the NIR part of HFF observation on A2744 has been completed, with the optical part currently underway, there is already a large number of multiple images identified, sufficient for our non-parametric method to be applied.

Previous lens models and claimed high-z galaxies recently detected using the HFF data behind A2744 are summarised in section \ref{sec:previous}. 
We construct our own filtered colour images in the optical and NIR and describe our photometric redshift measurements in section \ref{sec:colour}. 
In section \ref{sec:model} we describe our methodology of constructing a free-form lens model. 
We define "geometric redshifts" in section \ref{sec:geoz}, based on the distance-redshift scaling for multiply-lensed galaxies. 
In section \ref{sec:individual} we describe each of the 18 lensed image systems in turn, with relensed images, geometric and photometric redshift comparisons. 
Our main results are described in section \ref{sec:results}, with conclusions and discussion is section \ref{sec:discussion}. 
Standard cosmological parameters are adopted:
$\Omega_M=0.3, \Omega_\Lambda=0.7$ and $h=H_0/100\,{\rm km\,s^{-1}\,Mpc^{-1}}=0.7$.

\section{Previous models and lensed high-z detections}\label{sec:previous}
  
As part of the HFF project, several lensing models of A2744 have been offered in the MAST archive\footnote{http://archive.stsci.edu/prepds/frontier/lensmodels/}.
The models differ considerably in terms of their magnification maps, but all are based on a similar set of multiple images, most of which were first reported by \citet{merten11} and uncovered by the method of \cite{broadhurst05, zitrin09a}. 
This semi-parametric method has proven to be a most effective tool for predicting the location of counter images, based on the assumption that the distribution of dark matter can be approximated by distribution of member galaxies. 
In this method, the masses of member galaxies are scaled by their luminosity and co-added, and a low order smoothing applied with free coefficients to allow for some flexibility in modelling the general cluster component. 
Depending on the size of the lens and the complexity of the mass distribution, the typical positional uncertainty in predicting counter images using this method is in the range of 1-5".

Since the aforementioned models were made available, several new spectroscopic redshifts of lensed images have been reported by \citet{richard14, johnson14}, which we make use of in our work described below, as listed in table \ref{table1}. 
Additional VLT and Subaru images \citep{cypriano04,okabe08,okabe10,okabe10b} have also used to constrain these models. 
Comparisons have been made between these model makers and a joint effort to understand and quantify systematic errors and other uncertainties, and hence the performance of reconstruction tools were assessed with simulations of cluster observations \citep{meneghetti08, meneghetti10}. 

Recently, \citet{zheng14} reported 18 candidate Lyman-break galaxies (LBG) at $z\gtrsim$ 7 in the field of A2744. 
The faintest sources detected are around AB magnitude 28.5. 
A high quality sample of 24 objects with ``secure'' photometric redshifts greater than 7 is claimed, each with probability ($<1\%$) of being at low redshift. 
Another object at $z=8$ has recently been reported by \citet{laporte14}, and is consistent with the redshift estimated by \citet{zheng14}. 
An early publication by \citet{atek14} has claimed 6 sets of multiple images with photometric redshifts in the range $5<z<7$. 
As we will show, three of these sets need correcting, another seems spurious, and two others we complete with additional multiple images. 
\citet{johnson14} revised their parametric model with new Gemini spectroscopy for two of the multiply lensed systems of \cite{merten11}. 
\citet{zitrin14} discovered a double image pair of a z$\simeq$9.83 galaxy, becoming the multiply-lensed system with the highest redshift in the field of A2744. 
\citet{jauzac14} claimed $\sim$150 multiple images using the complete HFF data.

\section{Colour Images and Photometric Redshifts}\label{sec:colour}


We retrieved the drizzled science images from the online MAST HFF data archive. 
The individually reduced images are produced by \textit{Mosaicdrizzle} \citep{koekemoer13} and publicly available\footnote{http://archive.stsci.edu/pub/hlsp/frontier/abell2744/images/hst/v1.0/}. 
These deep images can probe faint galaxies to magnitudes of $\sim28-29$, not to mention the magnification due to cluster lensing, which could increase the limiting magnitude to $\sim31$. 
Our analysis uses the complete HFF data of A2744. 
We refer the details of the observations to \citet{zheng12} and the official website of HFF\footnote{http://www.stsci.edu/hst/campaigns/frontier-fields/HST-Survey}. 

The identification of multiple images is made difficult for counter images that are buried in the light of bright member galaxies and when significantly contaminated by diffuse light in the cluster core. 
We produce colour images after processing the seven ACS+WFC3 bands in order to (i) reduce the brightness from bright member galaxies and/or, cluster core, and (ii) increase the signal to noise ratio of faint distant objects. 

\begin{figure}
\includegraphics[width=85mm]{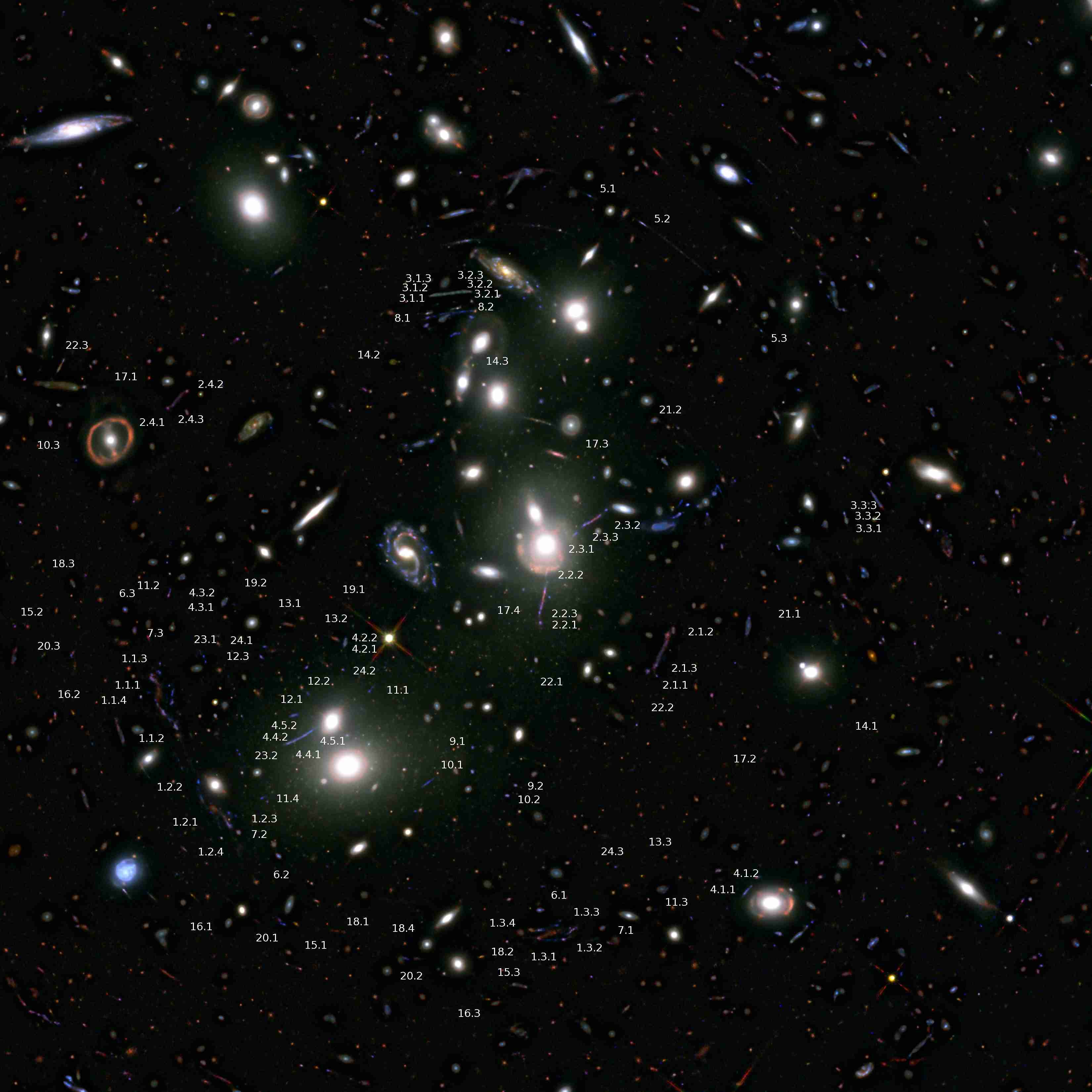}
\caption{
ID's of multiple-image systems overlaid on our NIR constructed colour image to enhance the visibility of faint red images at our highest redshifts. 
Note that in the process of generating this image, the filtering of objects can lead to colour changes of central images affected by the intra-cluster light. 
Here we only show the central 1.65"$\times$1.65" region, beyond which no multiply lensed images are found. 
The full resolution version can be found electronically online. 
}
\label{combo_map}
\end{figure}

To increase the signal-to-noise ratio of faint objects, we smoothed the images in each passband with a Gaussian filter of FWHM ranging from 60 milliarcsecond for the IR bands to 100 milliarcsecond for the optical bands. 
This smoothing also helps to compensate for the difference in angular resolution between the ACS and WFC3 cameras, producing images more nearly matched in angular resolution. 
To reduce the glare of bright member galaxies, we applied a low-pass filter to the individual bands thus reducing the diffuse light around the centre. 
In this way, the surface brightness and internal structures of the objects falling in this region can be more reliably compared with possible counter images at larger radius. 
Finally, as shown in figure \ref{combo_map}, we adopted a combination of a power law plus a Gaussian colour scaling to the images that we found provided a dynamical range of colours that matches the relatively faint range of brightness of the majority of lensed images.

\begin{figure}
\includegraphics[width=85mm]{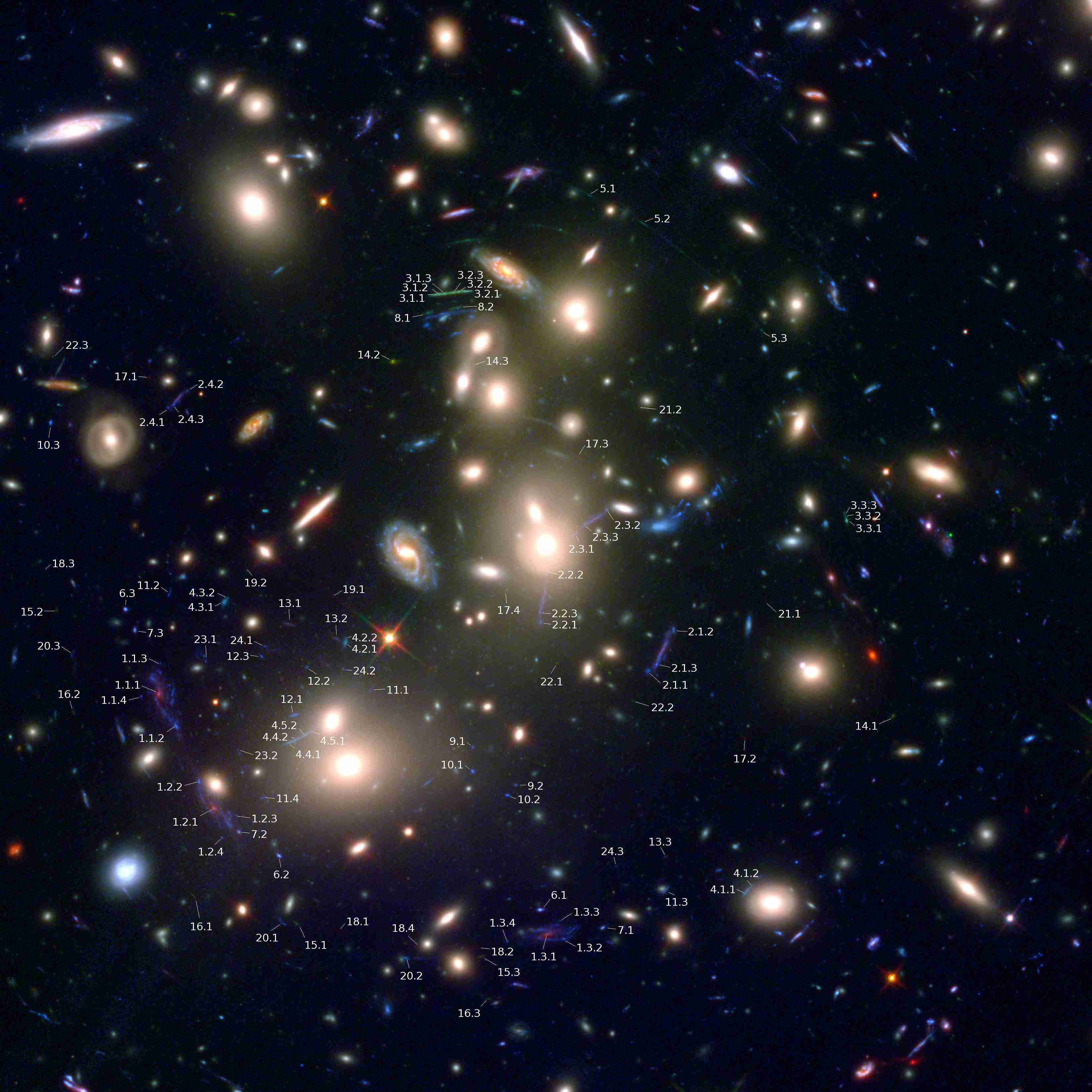}
\caption{
Same as figure \ref{combo_map}, except this is a simple RGB colour image which does not suffer from colour changes due to filtering. 
It is constructed from complete-HFF data, with F435W and F606W bands constituting the blue colour, F814W for green, and the NIR bands F105W, F125W, F140W, and F160W for red. 
The full resolution version can be found electronically online. }
\label{colour_map}
\end{figure}

The filtering process described above, however, causes colour changes in images close to the diffuse intra-cluster light. 
As a result, multiple images might be mistakenly associated if relying on colour information only. 
Therefore, we also generated a simple RGB image without subtraction of the smooth light component with the publicly available software \textit{Trilogy}\footnote{http://www.stsci.edu/~dcoe/trilogy/Intro.html}. 
This image is shown in figure \ref{colour_map}, and complements the previous high contrast image with its reliable colour.

Using the optimised colour images produced, we identify obvious sets of counter images for those sources with distinctive morphologies and unusual colours. 
These images are then used to generate an initial free-form model described in section \ref{sec:model} below. 
Note that in the case of highly-magnified long arcs, we include points along the length of the arc as our model is designed to make use of this additional information. 

Having generated an initial deflection field based on the obvious multiple systems, this model can then be used to help search for less obvious systems which are either less distinctive or for which counter images lie at locations that are not readily anticipated without the help of a model. 
For this purpose, free-form modelling is especially important because, unlike for more relaxed clusters, the lensing deflection field of A2744 is strongly perturbed by the on-going merging of sub-components as is evident from the multi-modal distribution of member galaxies and the observed spatial distribution of the cluster X-ray emission, and cannot therefore be readily foreseen.

\begin{figure*}
\centering
\includegraphics[width=160mm]{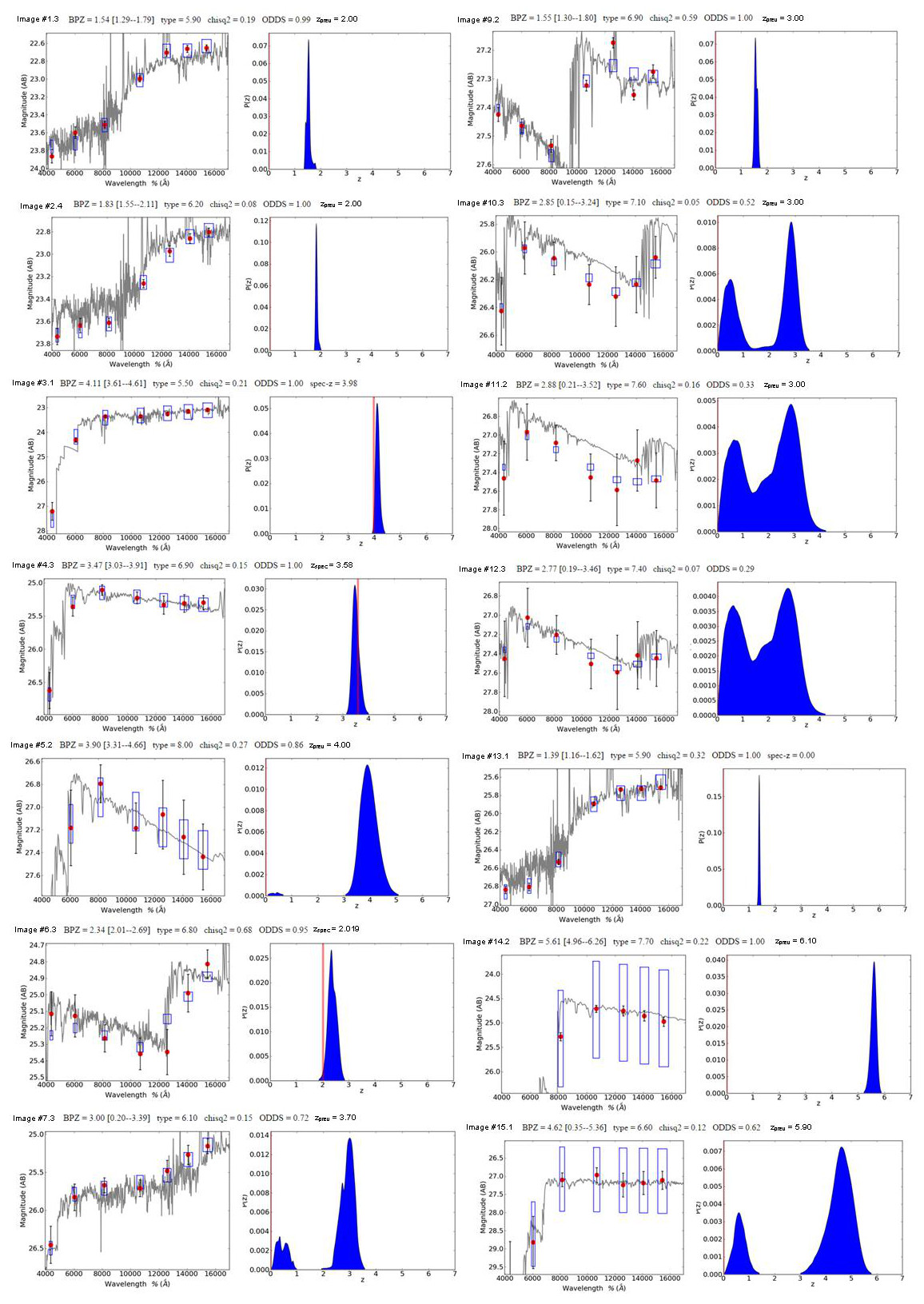}
\caption{
Best fitting template spectra compared to the SED's of selected lensed images and the probability distribution function in redshift space for systems 1 to 16. 
The red vertical line indicates the redshift reported by previous spectroscopic studies. 
Blue boxes are the model magnitudes for exact passband for the best fit SED, with the uncertainty of photometry. 
Our 2-sigma uncertainties and minimum chi-square are also quoted for each image. 
}
\label{bpz1}
\end{figure*}

\begin{figure*}
\centering
\includegraphics[width=160mm]{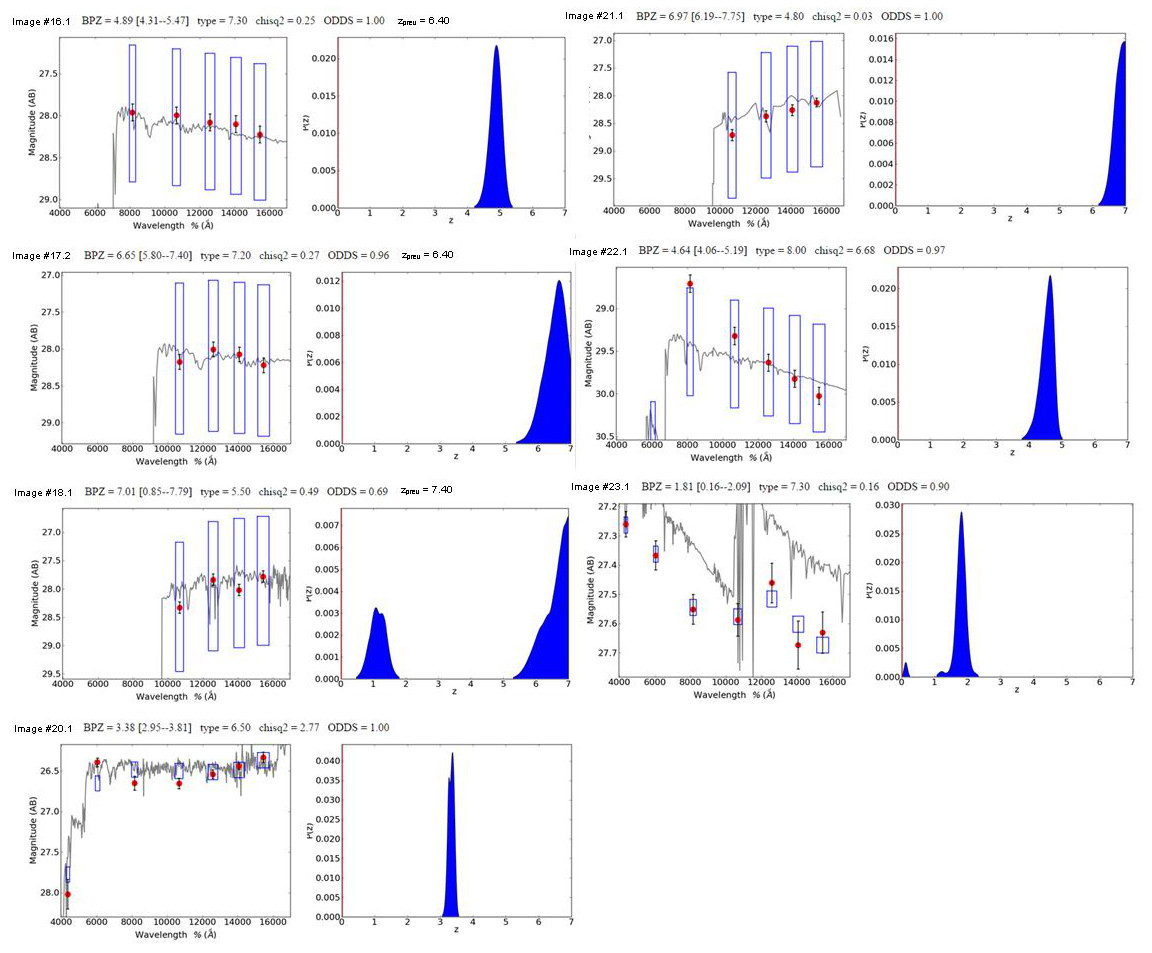}
\caption{
Same as figure \ref{bpz1} except here we show systems 17 to 23. 
}
\label{bpz2}
\end{figure*}

We also compare the spectral energy distributions of candidate multiple-image systems to ensure consistency, given the achromatic nature of gravitational lensing. 
Photometric redshifts for these images are estimated using the BPZ method \citep{benitez00, coe06}, for which the whole probability distribution is provided, as shown in figures \ref{bpz1} to \ref{bpz2}. 
The BPZ algorithm uses a template library which consists of model spectra of five elliptical galaxies, two spiral galaxies, and four starburst galaxies with moderately strong emission lines. 
Originally the templates were based on the P\'{E}GASE stellar population synthesis models \citep{fioc97}, but have been empirically corrected using a subset of sources from the FIREWORKS survey \citep{wuyts08}, for which photometric and spectroscopic redshifts can be compared. 
Flat priors on both galaxy type and redshift in the range $z=0-12$ are assumed here, as is standard practice. 
Apart from having similar SED, the positions of the images as well as their orientations must also be consistent with each other. For this purpose, we have developed a 'relensing' tool so that the pixels of any object can be delensed back to the source plane and then remapped onto the image plane to generate the expected appearance of counter-images. 
When relensing, we can make use of the photometric redshift information including its uncertainty, but this is not necessary and often ambiguous for very faint images. 
Instead, we created relensed images to cover a range of source distances defining ``loci'' along which we search for counter images. 
In this way, we can both reliably identify relensed counter-images and obtain purely geometric distance estimates for each set of multiple images, as described more fully in the next section.

Rather than using automated software to perform the photometry, we conducted our own photometry for most of the candidate lensed galaxies to construct their SEDs. 
The photometry for each lensed image is performed with tailored-made apertures and sky annuli. 
This refinement is necessary due to the highly distorted shapes of most of the candidate lensed galaxies, making it difficult for automated software to capture. 
It is also necessary for lensed images that appear close to member galaxies, where they are buried in bright intracluster light and could not be detected by the automated software. 
The size of each source aperture extends to where the source intensity falls to the noise level in the F160W band. 
The corresponding sky annulus is separated from the source aperture boundary by $\sim$0.1", and contains as many pixels as in the source aperture. 
The fits to SEDs and the resultant probability distribution function in redshift space along with relevant uncertainties are shown in figure \ref{bpz1} to \ref{bpz2} and summarised in Table \ref{table1}. 
In addition to the photometric redshifts we obtained from the SED fits, we also evaluated the validity of claims of multiple-image systems by checking whether the SED's are similar to each other. 
Unless a system has a reported spectroscopic redshift, the photometric redshift of each system with the best fitted SED is chosen to be the input redshift when constructing the lens model, as described in section \ref{sec:individual}.

\section{Free-Form Lensing Model}\label{sec:model}

The free form lensing method developed by \citet{diego05a} is a grid-based iterative method that can be constrained by both strong and weak lensing information, including sets of individual pixels subtended by resolved arcs in the case of strong lensing. 
We have recently demonstrated that this methods can be significantly improved by the addition of observed member galaxy deflections (Weak and Strong Lensing Analysis Package plus member galaxies: WSLAP+, \citet{sendra14}) because typically one or more counter-images of each multiply-lensed system is either generated or significantly deflected by a local member galaxy. 
We have applied this method recently to the relaxed cluster A1689 \citep{diego14} and demonstrated that this combination of high and low frequency components can converge to meaningful solutions with sufficient accuracy to allow the detection of new counter-images for further constraining the lensing solution of A1689. 
Because this method is free-form, it should be especially useful for modelling the complex mass distributions (which cannot be readily foreseen) of the clusters chosen for the Hubble Frontier Fields program. 
Parameterised models are inherently less useful in this context because many additional and uncertain parameters having limited constraints must be introduced.

Here we outline briefly this new method, WSLAP+, for the mass reconstruction and refer the reader for details of its implementation and testing to our previous papers
\citep{diego05a, diego05b, diego07, ponente11, sendra14, diego14}. \\

\noindent Given the standard lens equation, 
\begin{equation} \beta = \theta -
\alpha(\theta,\Sigma(\theta)), 
\label{eq_lens} 
\end{equation} 
where $\theta$ is the observed angular position of the source, $\alpha$ is the deflection angle, $\Sigma(\theta)$ is the surface mass density of the cluster at the position $\theta$, and $\beta$ is the position of the background source, both the strong lensing and weak lensing observables can be expressed in terms of derivatives of the lensing potential
\begin{equation}
\label{2-dim_potential} 
\psi(\theta) = \frac{4 G D_{l}D_{ls}}{c^2 D_{s}} \int d^2\theta'
\Sigma(\theta')ln(|\theta - \theta'|), \label{eq_psi} 
\end{equation}
where $D_l$, $D_{ls}$ and $D_s$ are, respectively, the angular diameter distances to the lens, from the lens to the source, and from the observer to the source. 
The unknowns of the lensing problem are in general the surface mass density and the positions of the background sources. 
As shown in \citet{diego05a}, the lensing problem can be expressed as a system of linear equations that can be represented in a compact form,
\begin{equation}
\Theta = \Gamma X, 
\label{eq_lens_system} 
\end{equation} 
where the measured lensing observables are contained in the array $\Theta$ of dimension $N_{\Theta }=2N_{SL}$, the unknown surface mass density and source positions are in the array $X$ of dimension $N_X=N_c + N_g + 2N_s$, and the matrix $\Gamma$ is known (for a given grid configuration and initial galaxy deflection field, see below) and has dimension $N_{\Theta }\times N_X$.  
$N_{SL}$ is the number of strong lensing observables (each one contributing with two constraints, $x$, and $y$), and $N_c$ is the number of grid points (or cells) that we divide the field of view ($2.56"\times2.56"$) into, which equals to $32^{2}=1024$ in this case.  
$N_g$ is the number of deflection fields (from cluster members) that we consider.  
$N_s$ is the number of background sources (each contributes with two unknowns, $\beta_x$, and $\beta_y$, see \cite{sendra14} for details). 
The unknowns are found after minimizing a quadratic function that estimates the solution of the system of equations \ref{eq_lens_system}, with the constraint that the solution, $X$, must be positive, and that the convergence of the solution does not exceed the angular resolution of the data. 
This constraint is particularly important to avoid the unphysical situation where the masses associated to the galaxies are negative (which could otherwise provide a reasonable solution, from the formal mathematical point of view, to the system of linear equations \ref{eq_lens_system}). 
Imposing the constraint $X>0$ also helps in regularizing the solution as it avoids large negative and positive contiguous fluctuations. 
The minimization is carried out on the source plane, meaning that our algorithm iterates in search for a solution (mass distribution and source positions) such that the image pixels of each lensed system converges on the source plane. 
At each iteration, the algorithm searches for a new solution in the direction of steepest descent (in terms of chi-squared), which is also required to be orthogonal to that of the previous iteration. 

\begin{figure}
\includegraphics[width=85mm]{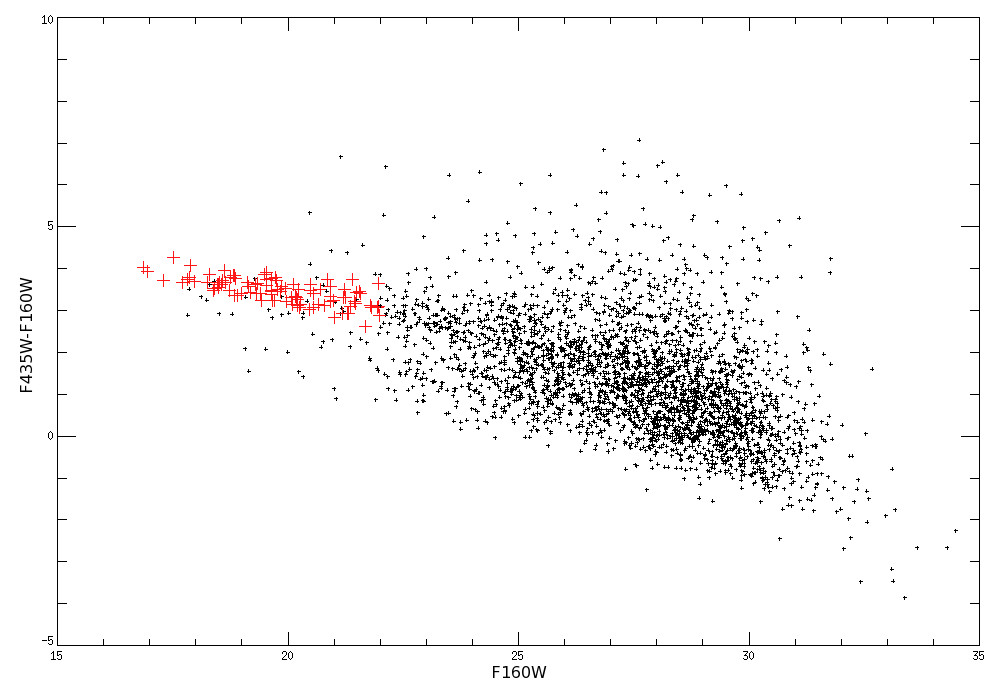}
\caption{
Colour-magnitude diagram of all detected light sources in A2744. 
The member galaxies lie along a clear red colour band at $m_{AB, F435W}-m_{AB, F160W}\sim4$. 
Member galaxies brighter than $m_{AB, F160W}$=22.0 and lie within the FOV of our analysis are selected to construct the fiducial deflection field and are marked as red crosses. 
There are in total 91 of them. 
}
\label{member}
\end{figure}

When sufficient constraints are available, the addition of the small deflections by member galaxies can help improve the mass determination. 
For our study we select the brightest elliptical galaxies (from the red sequence) in the cluster central region and associate to them a mass according to their luminosity. 
Member galaxies are selected to lie on the prominent colour-magnitude relation for early-type galaxies, by requiring their positions on the colour-magnitude diagram (F435W-F160W vs F160W) to be bounded by the empirical conditions y = -0.25*x + 8.0, y = -0.15*x + 7.0, and x = 22.0, as marked in large red crosses in figure \ref{member}. 
Several stars that satisfied the requirements are identified and removed manually. 
Finally, a total of 91 member galaxies were selected to construct the galaxy deflection field. 
From the H band (F160W) AB magnitudes, a mass-to-light ratio of 20 M$_{\odot}$/L$_{\odot}$ is initially assumed to construct the fiducial deflection field summed over the member galaxies, each having a truncated NFW profile (truncation radius equals scale radius times concentration parameter) with a scale radius linearly related to its FWHM in the NIR image. 
For our purpose, the exact choice of profile for member galaxies is not particularly important; what matters more is the normalisation. 
This normalization is the only free parameter of the fiducial deflection field, and is determined by our optimization procedure. 
In \cite{sendra14} we tested this addition to the method with simulated lensed images, but with the real galaxy members from A1689 to be as realistic as possible. 
We also found that a separate treatment of the BCG lensing amplitude was warranted, adding a second deflection field i.e $N_g=2$ (see definition of $N_g$ above) to be solved for. 
Here we follow the same procedure for A2744 incorporating member galaxies and the BCGs separately. 
We also find significant improvements in the residual by leaving free the amplitude of bright galaxies that significantly perturb nearby lensed images. 
In total, as shown in figure \ref{member_mass}, we decomposed the fiducial deflection field into 10 components, comprising the central cD galaxy, the cD galaxy in the SE and its companion, several other groups of galaxy where there are multiply-lensed images nearby, and the rest of the member galaxies. 

We estimate the uncertainties of the mass distribution, source positions, and magnification map using the ensemble of solutions we find compatible with the data. 
The range of models can be explored using the variance of the set of solution each covering a range of initial parameters including grid cell masses, and normalization of fiducial deflection fields. 
We performed in total 9 minimizations with the same constraints, and initial masses in each grid cell ranging from $1\times10^{11} M_{\odot}$ to $4\times10^{11} M_{\odot}$, and normalization of fiducial deflection fields from 0.5 to 2.0, where a value of 1.0 corresponds to M/L=20 $M_{\odot}/L_{\odot}$.

\begin{figure}\label{member_mass}
\includegraphics[width=85mm]{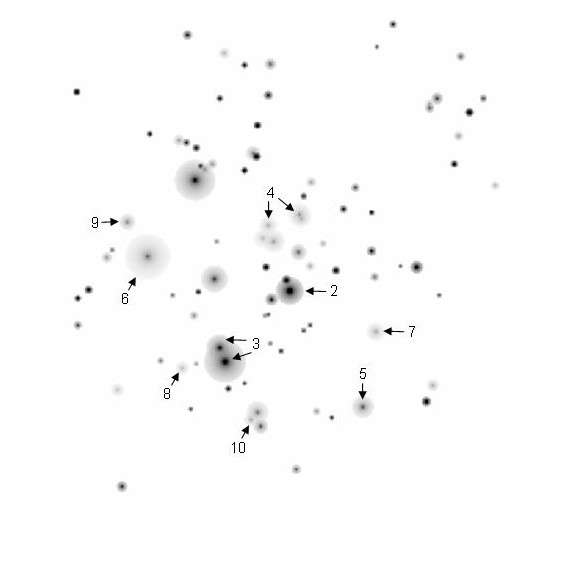}
\caption{
The 91 member galaxies selected to construct the fiducial deflection field. 
Circularly symmetric NFW haloes are assigned to each member galaxy with mass proportional to the F160W flux and scale radius proportional to the FWHM. 
To accommodate the intrinsic variation of mass-to-light ratio, and thus the degree of perturbation on nearby lensed images, we divide the member galaxies into 10 groups and assign each an independent M/L. 
Group 1 is not marked here as it corresponds to the remaining galaxies. 
The field of view is $2.56"\times2.56"$. 
}
\label{member_mass}
\end{figure}

\section{Geometric Redshifts}\label{sec:geoz}

Here we describe how we obtain source distances geometrically from the relative angles between sets of counter-images. 
The derived distances provide a very welcome check on redshifts derived photometrically, particularly at high redshifts where images are generally noisy and may be detected in only the longest-wavelength passband. 
For this purpose an accurate lens model is required, based on many sets of multiply-lensed images and ideally sampling a wide range of source distances so that the gradient of the mass profile can be constrained. 
The reduced deflection field scales with increasing source distance behind a given lens so that the separations in angle between images of the same system are larger for higher redshift sources. 
Distances derived this way can then be converted via cosmological parameters to source redshifts and compared with independently derived photometric redshifts. 
This method has been established using the large lensing cluster A1689, where geometric distances provided a consistency check of the lens model \citep{broadhurst05, limousin07, diego14}.

A lens model is a deflection field, $\vec{\alpha}_L(\vec{\theta})$, that expresses the angle through which light is bent at the lens plane. 
An observer sees a reduced angle scaled by a ratio involving lens and source distances:

\begin{equation}
\vec{\alpha}(\vec{\theta})=d_{ls}(z)/d_{s}(z) \vec{\alpha}_{L}(\vec{\theta})
\end{equation}

\noindent As mentioned above, the angles between the unlensed source and the lensed images increases with source distance for a given lens. 
This dependence means that the locations of a given set of multiple images will meet most closely in the source plane at a preferred source distance. 
In principle, we can only determine relative distances this way because the absolute value of $\vec{\alpha}_L(\vec{\theta})$ cannot be determined independently of lensing. 
By normalizing the model deflection field using a spectroscopic redshift measured for any one of the multiply-lensed systems, however, relative distances can be converted to absolute distances a given cosmological model. 
In other words, what we actually determine, for the $k^{th}$ set of multiple images for a given lens, is the ratio of lensing distances:

\begin{equation}
f_k(z)={d_{ls_k}(z) \over d_{s_k}(z)} / {d_{l{s_o}}(z_o) \over d_{s_o}(z_o)}
\end{equation}

\noindent In the case of A2744 we make use of the secure spectroscopic redshift of $z=3.580$ for system 4 to provide our normalization, $z_o$ (see Table \ref{table1}), measured by \citet{richard14}.

\section{Results}

\subsection{Individual multiply-lensed systems}\label{sec:individual}

\begin{figure*}
\centering
\includegraphics[width=150mm]{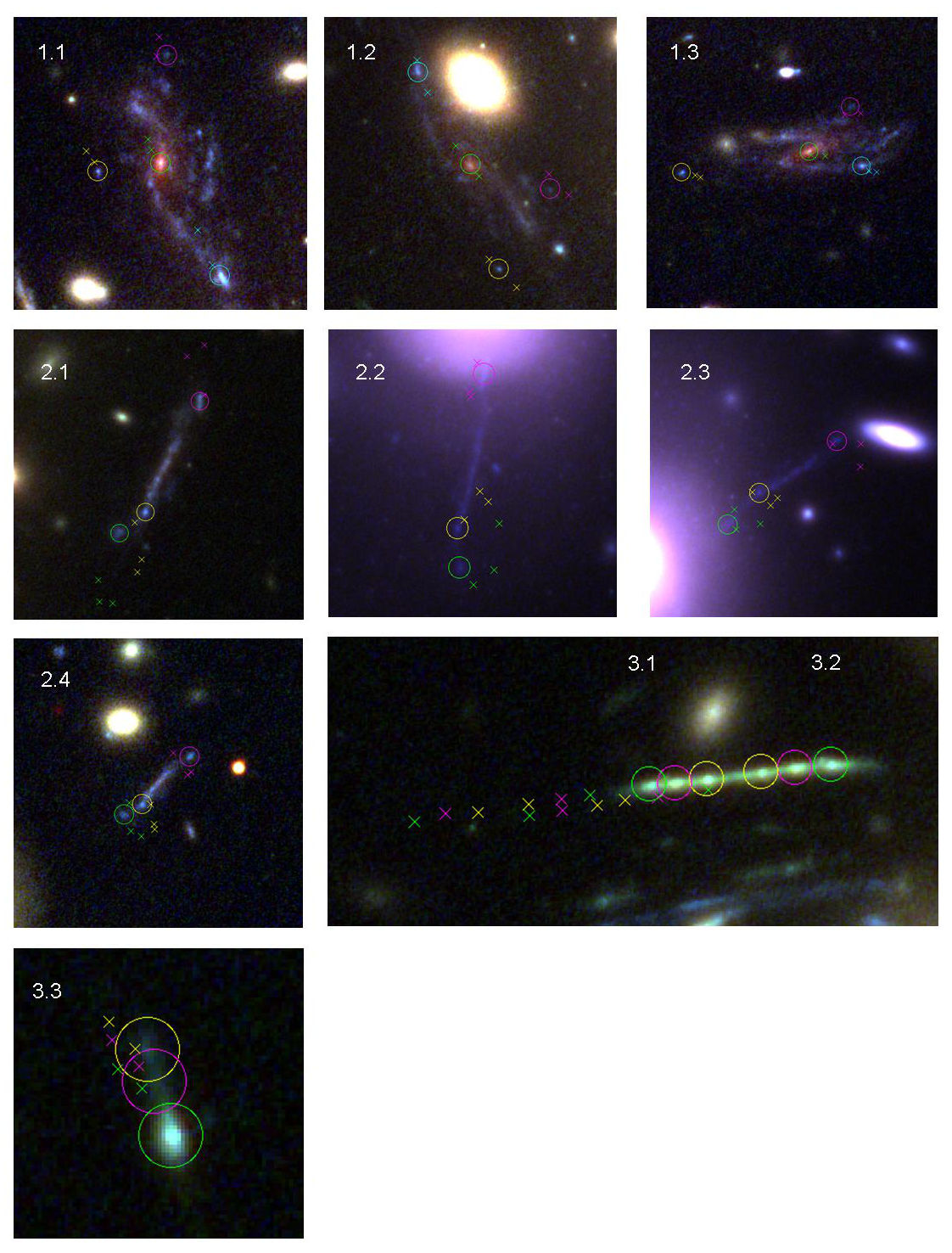}
\caption{
Image positions predicted by our lens model (marked as x's) for systems 1 to 3. 
Each marker is obtained by delensing and relensing the centroid of the corresponding image with photometric/spectroscopic redshift as input. 
The observed centroids are marked by circles, each with a diameter of 0.5". 
For well-resolved images we are able to compare the relensed positions of distinctive internal structures, 
which are distinguished from each other by different colour schemes. 
}
\label{stamps1}
\end{figure*}

\begin{figure*}
\centering
\includegraphics[width=150mm]{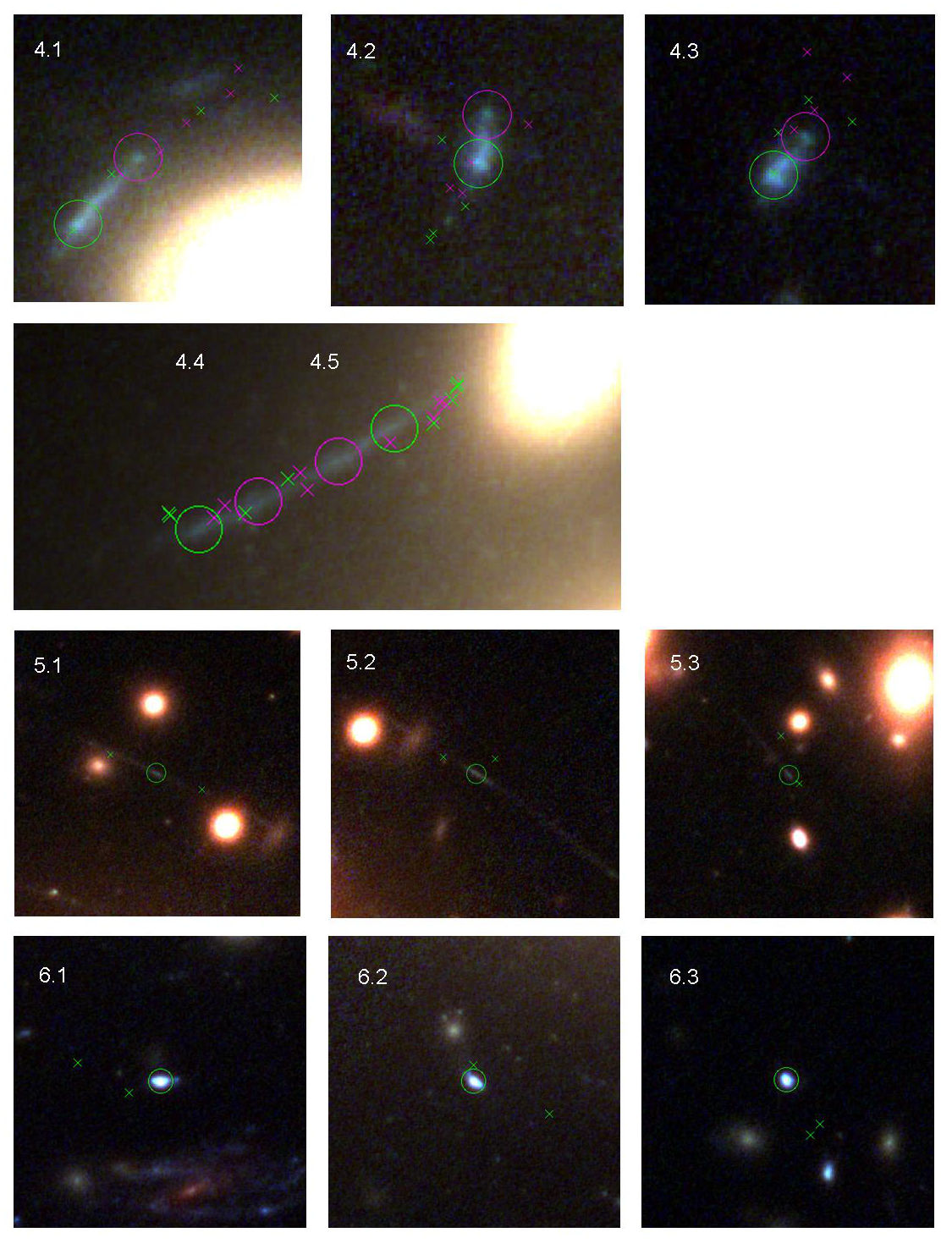}
\caption{
Same as figure \ref{stamps1} except here we show systems 4 to 6.
}
\label{stamps2}
\end{figure*}

\begin{figure*}
\centering
\includegraphics[width=150mm]{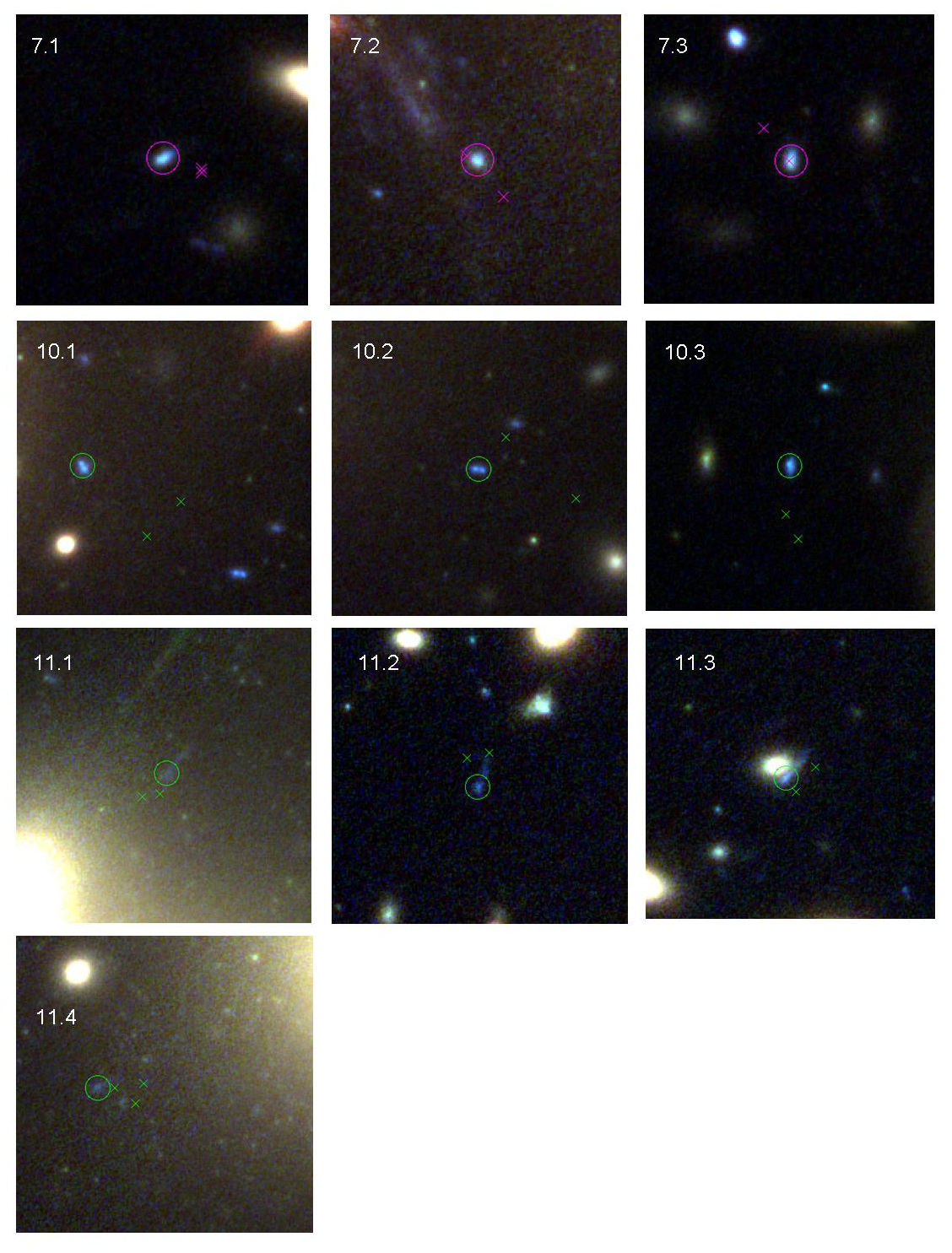}
\caption{
Same as figure \ref{stamps1} except here we show systems 7 to 11.
Systems 8 and 9 are not used in constructing our lens model because their photometric redshifts are not confidently measured.
}
\label{stamps3}
\end{figure*}

\begin{figure*}
\centering
\includegraphics[width=150mm]{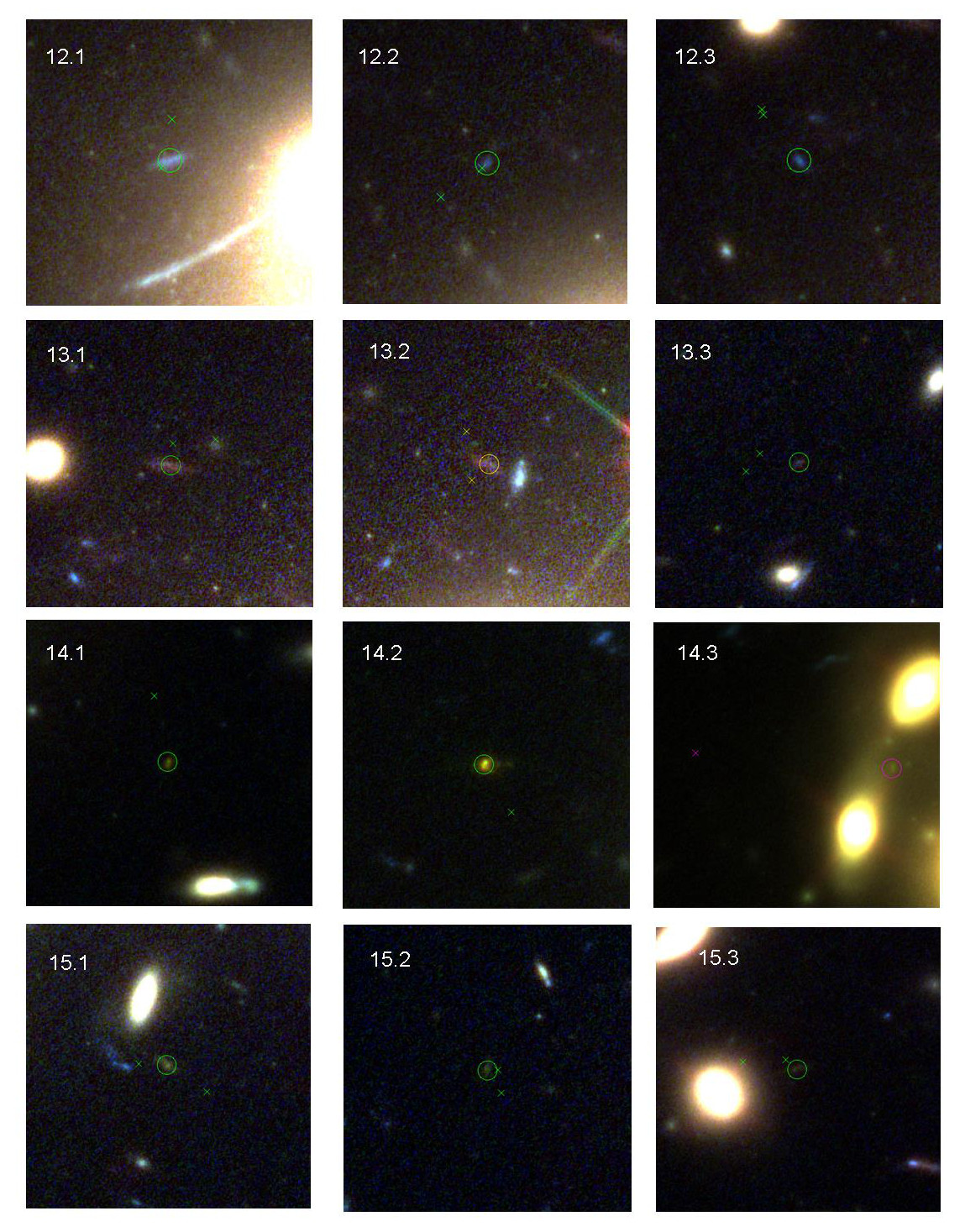}
\caption{
Same as figure \ref{stamps1} except here we show systems 12 to 15. }
\label{stamps4}
\end{figure*}

\begin{figure*}
\centering
\includegraphics[width=150mm]{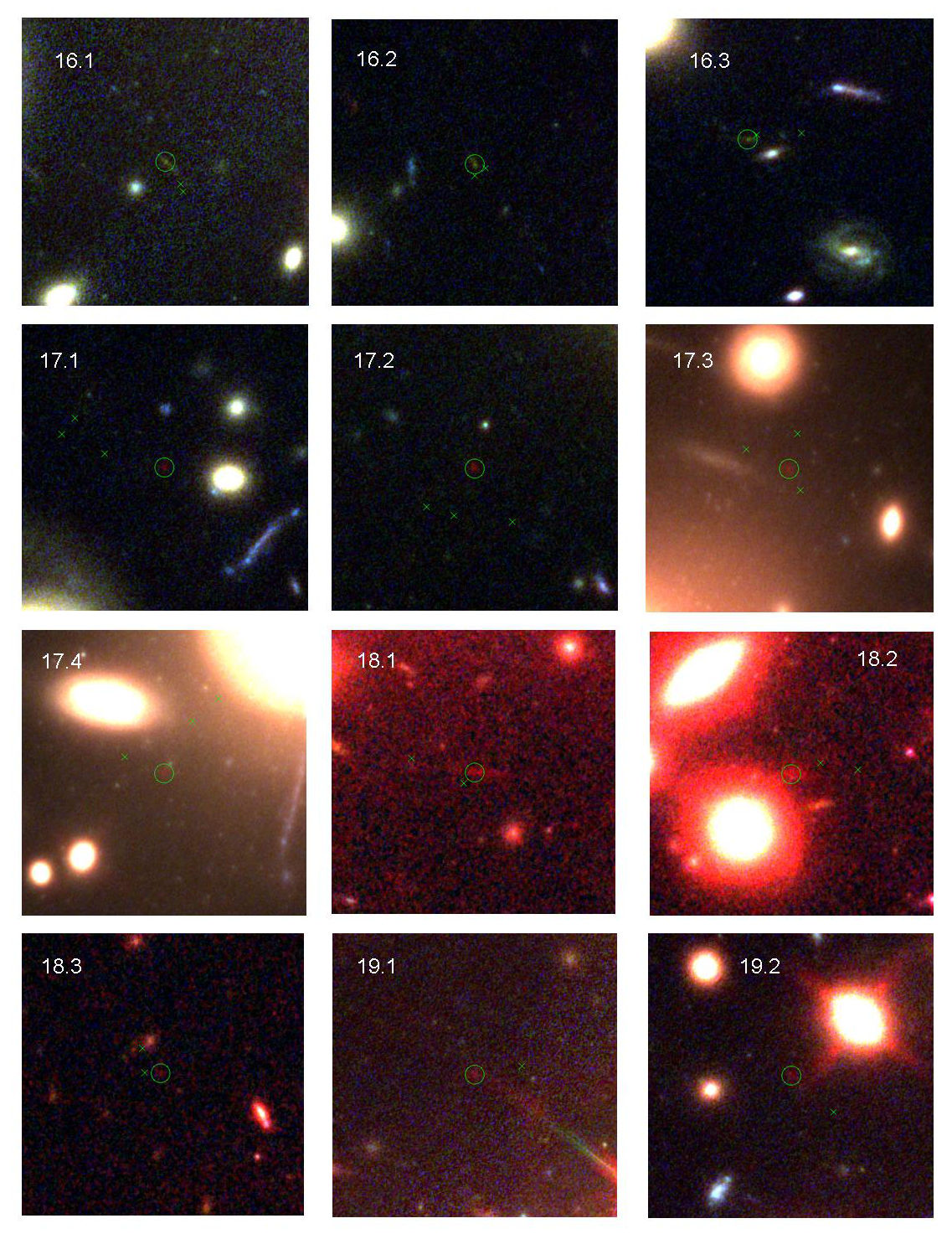}
\caption{
Same as figure \ref{stamps1} except here we show systems 16 to 19. 
Stamps of systems 18 and 19 are displayed in an image which we enhance the red colour to better view high redshift images. 
}
\label{stamps5}
\end{figure*}

\begin{figure*}
\centering
\includegraphics[width=140mm]{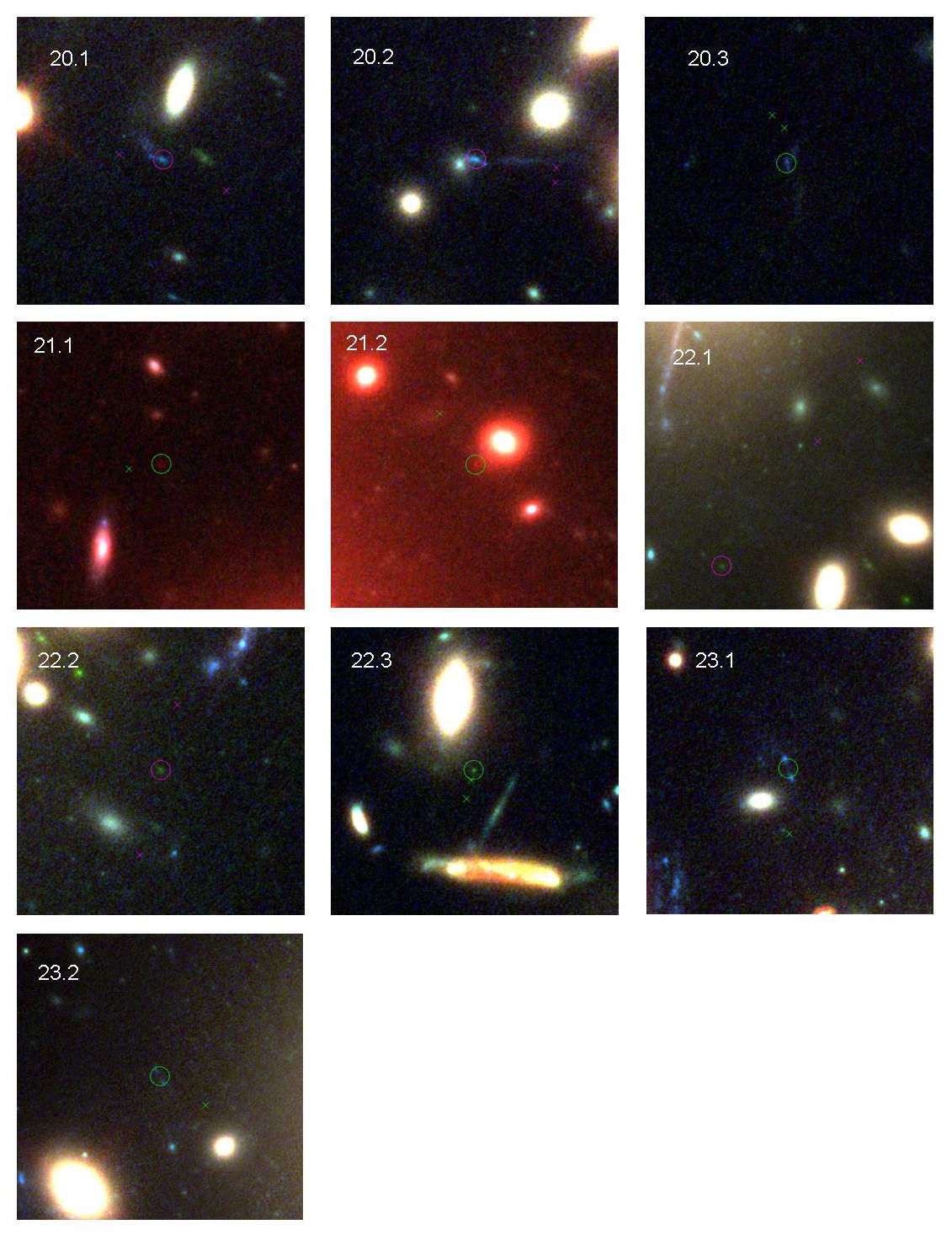}
\caption{
Same as figure \ref{stamps1} except here we show systems 20 to 23. 
}
\label{stamps6}
\end{figure*}

\begin{figure}
\centering
\includegraphics[width=85mm]{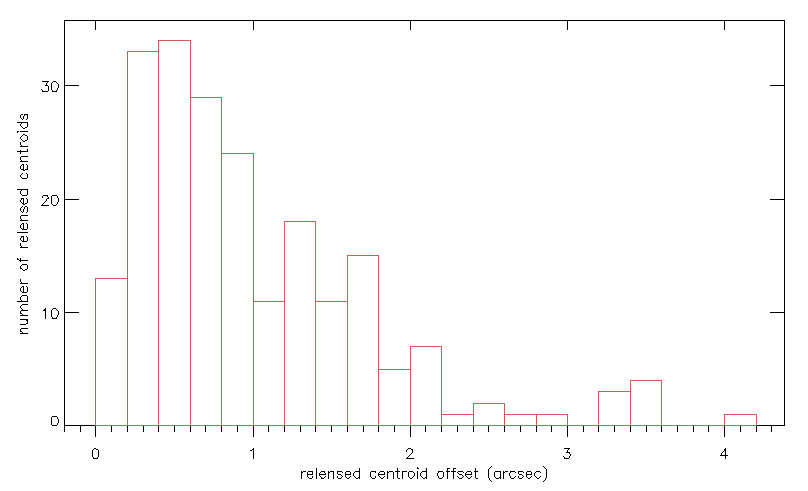}
\caption{
A histogram of offsets between predicted and observed image centroids. 
The offsets are typically around 0.2" - 0.4", with a mean of 1.0", and an RMS error of 1.25". 
Several images have offsets larger than 2" because they are located around the critical curves, making their predicted positions very sensitive to small changes in the deflection field. 
}
\label{histogram}
\end{figure}

Here, we describe the multiply-lensed systems used to derive the lens model for A2744, as well as those not used to constrain our lens model for reasons that we will explain. The ID number, positions, redshifts in table \ref{table1}, and magnification, photometry on these systems are tabulated in table \ref{table2}. 
Systems 1 to 11 were previously identified by the method of \citet{zitrin09a} and listed in \citet{merten11}, while systems 12 to 17 were identified recently by \citet{atek14}. 
We identified system 18, which is included in \citet{zheng14} prior to this work. 
System 19 was identified by \citet{zitrin14} and geometrically supported by our model. 
We identified 4 new systems, number 20 to 23, with the full HFF data which completed recently. 
These systems are also independently identified by \citet{jauzac14}. 
More system candidates are being studied in detail and are not included in table \ref{table1}. 
The accuracy of our identifications is demonstrated by delensing and relensing the well resolved images that have notable elongation/distortion and distinct internal structures. 
For these, as well as other multiply-lensed images that are not well-resolved, we also indicate the centroids of each predicted counter images in figures \ref{stamps1} to \ref{stamps6}. 
The distribution of such positional offsets between predicted and observed images are shown in figure \ref{histogram}. 
It has a modal value of 0.4'', a mean of 1.0'', and an rms of 1.2''. 
Some outliers correspond to those lying close to the critical curves where predicted positions become very model/redshift sensitive.

\begin{figure*}
\centering
\includegraphics[width=150mm]{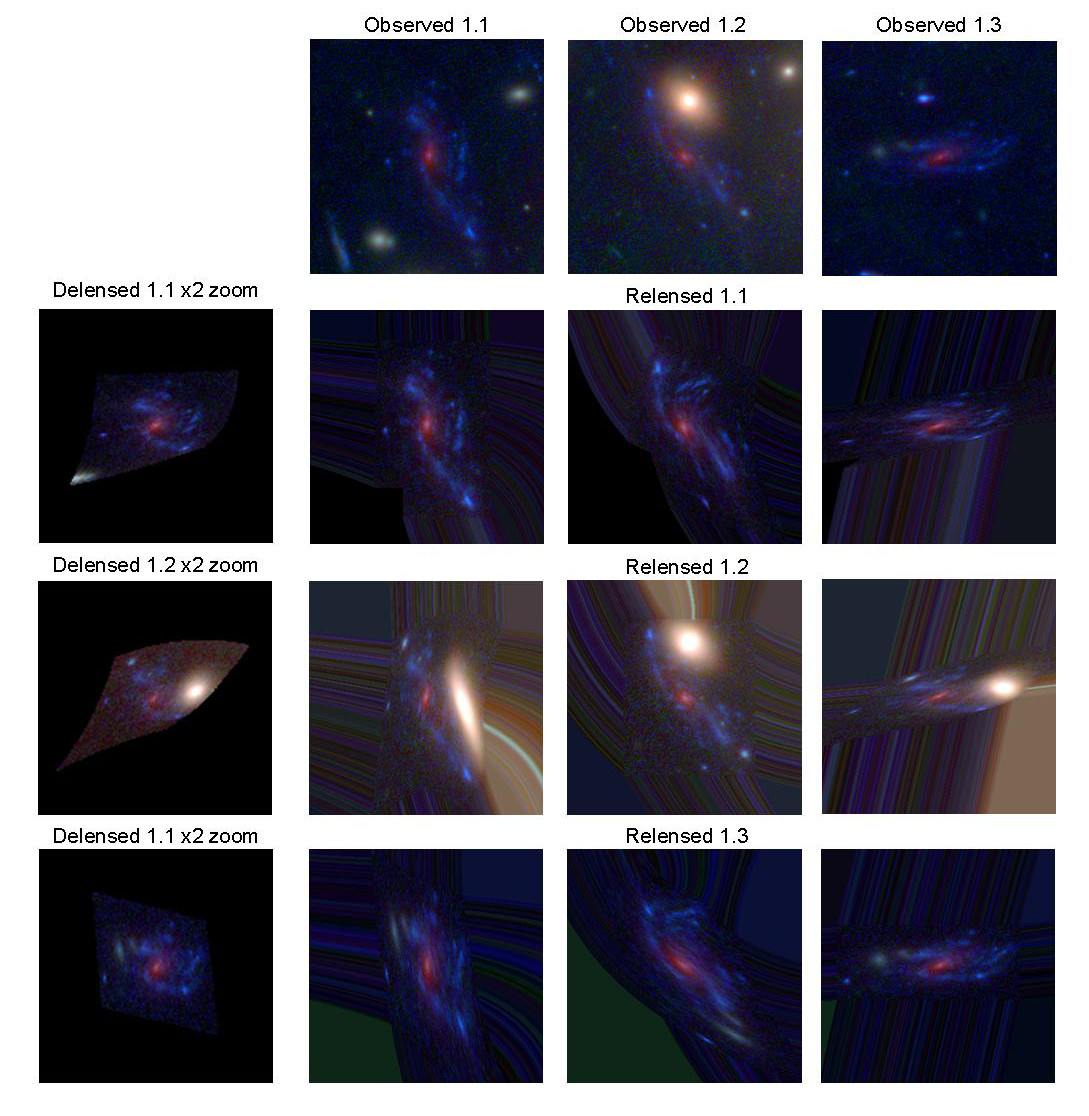}
\caption{
Delensing and relensing system 1. 
The top row shows the three observed images of system 1. 
The size of each stamp is 6.6" x 6.6". 
The left column displays the delensed images on the source plane enlarged by a factor of 2 in each dimension. 
The remaining panels show the images reproduced by our lens model, using all of the multiple images, to demonstrate the self-consistency of our lens solution. Note that released images of a given image must be identically equal within the noise and are shown only for completeness.
}
\label{system1}
\end{figure*}

System 1.  
This is a triply lensed spiral galaxy, which is the largest multiply-lensed system.  
We compute the photometric redshift of image 1.3, which is the one that is least contaminated by cluster light, thus presumably the one that yields the most accurate photometric redshift. 
The probability distribution peaks at z$\sim$1.5 with a secondary peak at z$\sim$1.8. 
A spectroscopic study by \citet{johnson14}, however, failed to detect any emission lines on this image between 440nm and 980nm, thus placing a lower limit of z>1.6. 
As a result, we chose the second likely photometric redshift z=1.80 as input. 
Here we also provide our own model reconstruction of the images, using the pixels of each image in turn to generate the other two images for comparison with the observed images, as shown in figure \ref{system1}. 
This reconstruction is done by solving for the lens equation, and then delensing followed by relensing of the flux.

\begin{figure*}
\centering
\includegraphics[width=150mm]{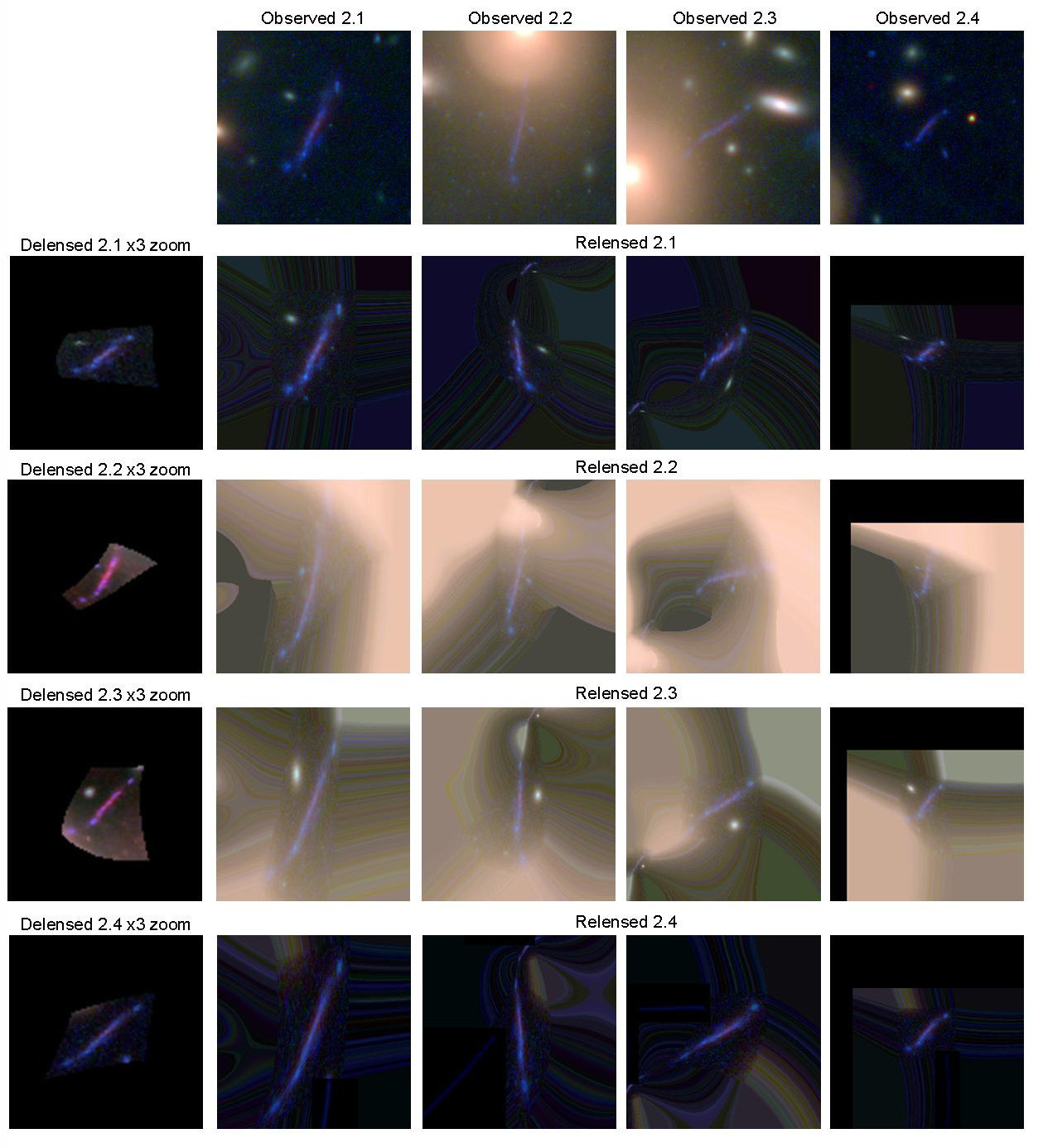}
\caption{
Delensing and relensing system 2. 
The top row shows the four observed images of system 2. 
The size of each stamp is 6.6" x 6.6". 
The two radial arcs are crop-outs from figure 2, which enhanced contrast for images close to the diffuse galactic light. 
The left column displays the delensed images on the source plane enlarged by a factor of 4 times in each dimension.
The remaining panels show the images reproduced by our lens model. Notice that the long radial arcs are well produced by our lens model and this has required a separate adjustment of the BCG mass profile within our model, as described in section \ref{sec:model}. 
Hence, these long arcs will provide detailed constraints on the mass distribution of the BCG in forthcoming work. 
}
\label{system2}
\end{figure*}

System 2. 
This is a four-image system where two long radial arcs are focused on the central BCG and two tangential arcs lie further out in radius, outside the tangential critical curve. 
As seen in figure \ref{system2} we securely demonstrate that these four images are related by successively delensing and relensing to generate predictions for the other three. 
We can see that the agreement between the predicted images and the observations is in general very good in terms of the predicted centroids. 
There is a small difference in the orientation of the central radially directed images, indicating the DM distribution of the central cD galaxy may be more complex than the circular symmetry assumed in our model. 
This discrepancy and the presence of these long radial images motivate modelling the profile of the cD galaxy separately. 
A fuller discussion of the constraints achievable with this unusually detailed information will be elaborated more fully in a forthcoming work (Lam et al. in prep.). 
We use the tentative spectroscopic redshift of z=2.2 \citep{johnson14} as input, which is consistent with our measured photometric redshift of $z=1.86^{+0.73}_{-0.28}$. The geometric redshift we derive of z=2.22$\pm$0.03 is in good agreement with the input redshift, showing a high level of self-consistency.

\begin{figure*}
\centering
\includegraphics[width=110mm]{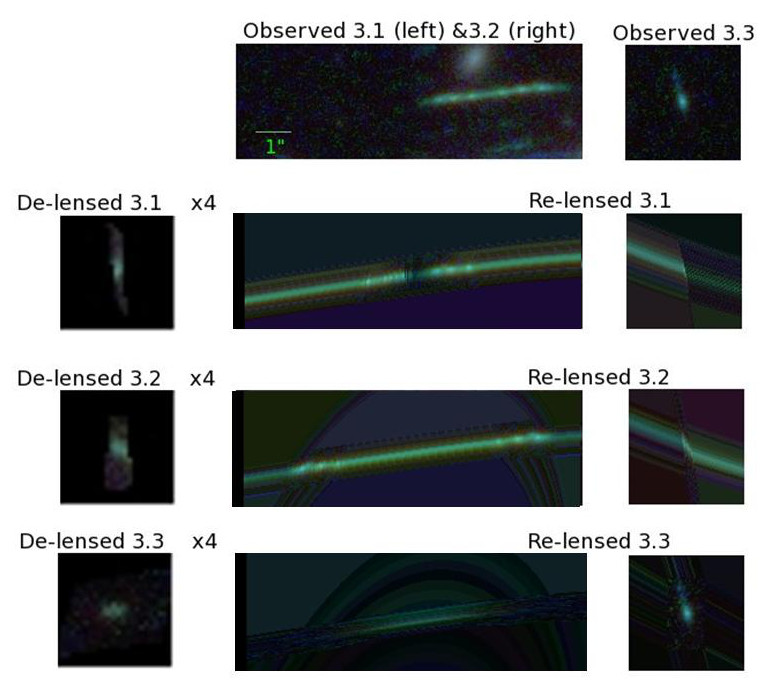}
\caption{
Delensing and relensing system 3. 
The top row shows the three observed images of system 3, including a close pair of highly magnified double images (3.1 and 3.2) that shows internal structures of three knots on each side of the critical curve. 
As mentioned in the text, the critical curve of our lens model did not pass through the middle of 3.1 and 3.2 as one would desire. 
Rather, the critical curve lies slightly to the East of 3.1 which makes the re-lensed images of 3.1 and 3.2 appear in unsatisfactory locations. 
Although re-lensing 3.1 and 3.2 produced third images elongated in the correct direction, their positions are far ($\sim$4") from the observed one. See figure \ref{stamps1} for the predicted and observed locations. 
}
\label{system3}
\end{figure*}

\begin{figure*}
\centering
\includegraphics[width=110mm]{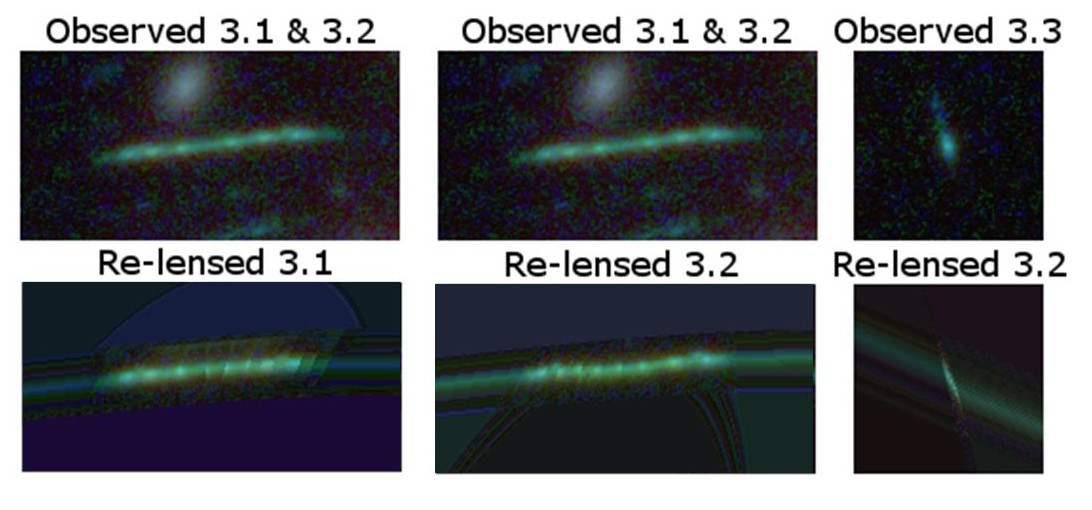}
\caption{
Delensing and relensing system 3 with a deflection field of amplitude reduced by $\sim7\%$. 
The top row shows the three observed images of system 3, and the bottom row shows the de-re-lensed images. 
Compared to figure \ref{system3}, one can see that drastically different results of de-lensing and re-lensing near the critical curve can arise from small changes in the deflection field. 
}
\label{system3a}
\end{figure*}

\begin{figure*}
\centering
\includegraphics[width=110mm]{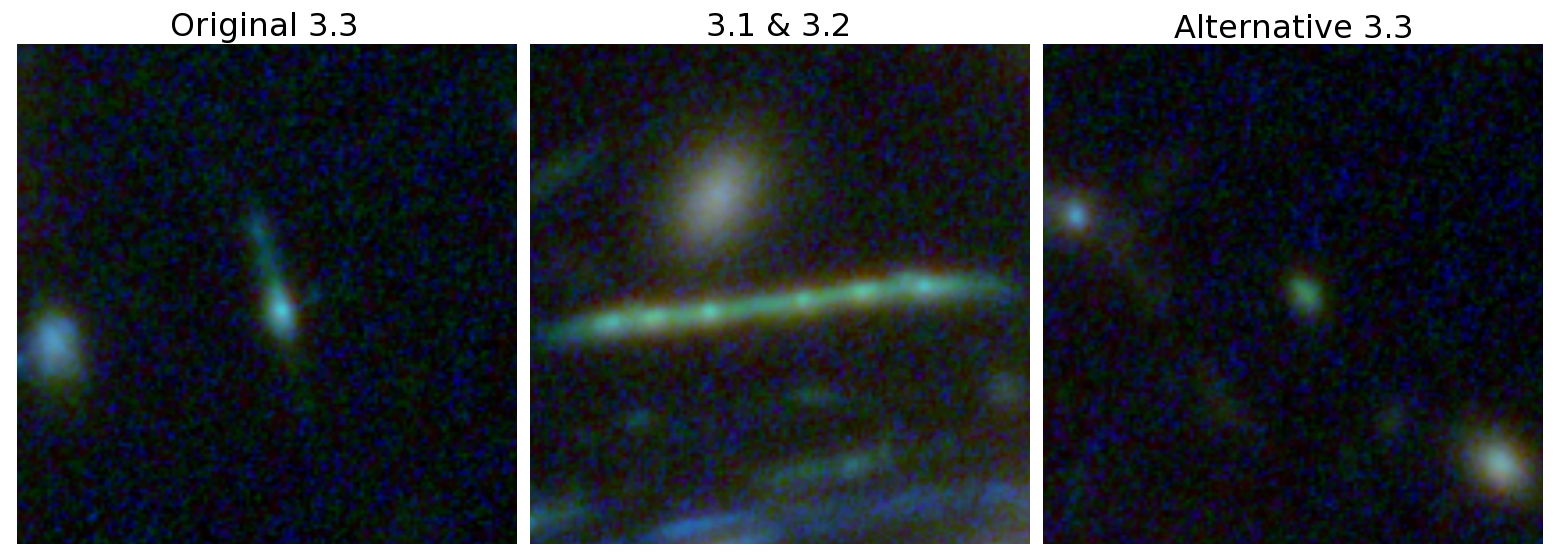}
\caption{
Close-up stamps of (left to right) 3.3 originally listed in \citet{merten11}, 3.1 and 3.2, and 3.3 alternatively proposed by us prior to the arrival of HFF optical data. 
With the complete HFF data, it is obvious that the colour of the original 3.3 resembles that of 3.1 and 3.2 better than the alternative 3.3 does. }
\label{system_3_colour}
\end{figure*}

System 3. 
This long arc, which has six knots, exhibits mirror symmetry about its centre, indicating that it comprises a pair of close images , labelled 3.1 and 3.2, straddling the tangential critical curve. 
The tangential critical curve of this ``caterpillar'' shaped object runs between images 3.1 and 3.2 showing reflection symmetry. 
Our non-parametric algorithm, however, tends to produce a lens model with the critical curve lying 1'' to the west of 3.1 rather than passing between 3.1 and 3.2; i.e., a difference in the expected position of the critical curve of 2''.  
As a result, additional images that are obviously not observed are predicted on the other side of the critical curve when we delens and relens 3.1 and 3.2, as shown in figure \ref{system3}. 
This is because the positions of lensed images appearing near the critical curve is very sensitive to slight changes in the deflection field, rather than because of wrongly identified multiple images. 
In figure \ref{system3a}, we show that a $\sim7\%$ reduction in the amplitude of the deflection field can produce the pair of mirror images 3.1 and 3.2 that closely resembles the observed images. 
We previously identified an alternative third image for this system (located at 00:14:18.39, -30:24:06.53) because its colour was indistinguishable from the original 3.3 (0:14:18.595, -30:23:58.42) listed in \citet{merten11}. 
As anticipated, the new optical HFF data settled this ambiguity. 
In figure \ref{system_3_colour} we show that the original 3.3 has a colour that is consistent with that of 3.1 and 3.2, while the alternative 3.3 does not. 
The geometric redshift that we derive for this system of z=4.19$\pm$0.27 is very close to that derived photometrically of z=4.11$\pm$0.50, and is consistent with the recently acquired spectroscopic redshift of z=3.98 \citep{johnson14}.

\begin{figure}
\centering
\includegraphics[width=85mm]{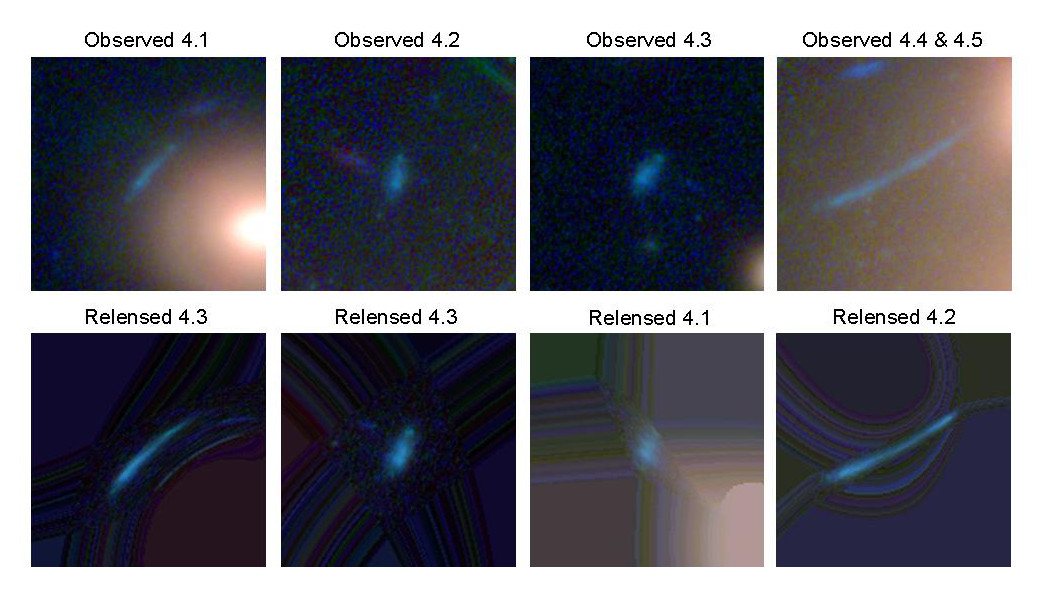}
\caption{
Delensing and relensing system 4. 
The top row shows the observed images of system 4 and for comparison the bottom row shows selected relensed images. 
The size of stamps is 3" x 3". 
}
\label{system4}
\end{figure}

System 4. 
This system comprises a double-image pair straddling the critical curve, along with a fainter counter image at a large radius from the BCG. 
In addition, a long radial arc was identified crossing a radial critical curve centred on a bright cluster member to the South. 
We show that all these images belong to the same source by successively delensing and relensing each image, selected examples of which are shown in figure \ref{system4}. 
A number of re-lensed cases are not useful to show where the observed image is of low contrast against the extended light of a member galaxy. 
The SED of 4.3 is compatible with a photometric redshift of z=3.47$\pm$0.44, which is consistent with the spectroscopic redshift of z=3.580 \citep{richard14, johnson14}.

System 5. 
A long tangential arc in the North consisting of 3 multiple images. 
The highly elongated nature of this image requires tailored aperture photometry. 
We chose the photometric redshift of image 5.2 (z=3.90$^{+0.76}_{-0.59}$) as input because it is the brightest of the system.

System 6.  
A high surface brightness triple system. 
We determine a photometric redshift of z=2.34$\pm$0.33 and geometric redshift of z=2.16$\pm$0.10, both in good agreement with the spectroscopic redshift of z=2.019 \citep{richard14, johnson14}. 
This object accurately falls on our distance-redshift relation as its spectroscopic redshift was used as input for our lens model reconstruction. 
We use this system as our model normalization as described in section \ref{sec:geoz}.

System 7. 
The derived photometric redshift of 7.3 is $z=3.00^{+0.39}_{-2.80}$. 
The large uncertainty towards the low redshift is due to a bimodal probability distribution. 
Since the spatial configuration of this triple system is similar to system 6, they should have comparable redshifts, thus ruling out the possibility of a being a low-redshift interloper.

System 8. 
This system has previously been identified as a triply-lensed system, but it was not used to constrain our lens model for reasons that we will explain. 
We are confident that system 8 comprises a close pair of very faint images labelled 8.1 and 8.2. 
A third image, is tentatively claimed by \citet{merten11}. 
This system is not included into constructing our lens model because we lack confidence in the identification of the 3rd image, which lies in a crowded field where accurate photometry is difficult to perform.

System 9.  
A pair of faint images and a unidentified third image possibly at a large radius. 
We measure the photometric redshift of 9.2 to be $z=1.74^{+1.34}_{-1.62}$. 
The large uncertainty reflects a broad and complex probability distribution in redshift space. 
As we cannot confidently measure its redshift, system 9 is not included into constructing our model.

System 10. 
A triply-lensed system with a spatial configuration similar to system 9. 
We measure the photometric redshift $z=2.85^{+0.39}_{-2.70}$ with the third image, as it is far from the contamination of cluster light, in which 10.1 and 10.2 situate in.

\begin{figure*}
\centering
\includegraphics[width=150mm]{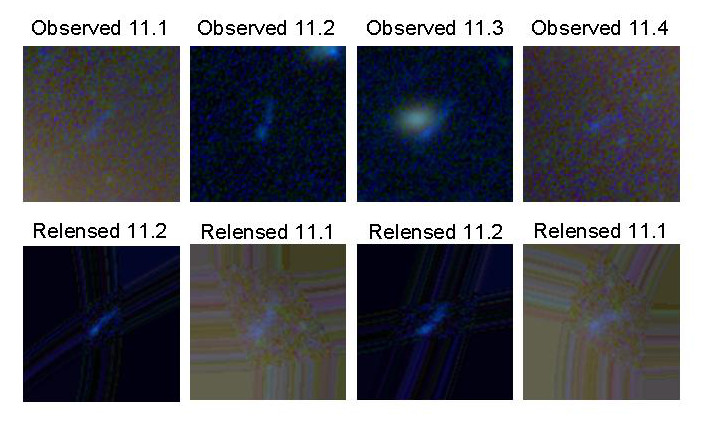}
\caption{
Delensing and relensing system 11. 
The top row shows the observed images of system 11 and for comparison the bottom row shows selected relensed images. 
}
\label{system11}
\end{figure*}

System 11. 
This is a faint bimodal object with four images. 
We infer a geometric redshift of z=2.76$\pm$0.02, which is in good agreement with the photometric redshift of $z=2.88^{+0.64}_{-2.67}$. 
Again, the large uncertainty towards the low-redshifts is due to the bimodal probability distribution. 
The fact that its multiple images are separated by large angles imply that it being a low-redshift interloper is improbable. 
We also relens the system as it is extended and has a double structure. 
As shown in figure \ref{system11}, we find good agreement in tens of the observed images and those we predict.

System 12. 
It consists of 3 multiple images each located not far away from one another. 
This less common configuration arises because they are lensed near a Y-shaped junction of critical curves. 
Its measured photometric redshift of $z=2.77^{+0.69}_{-2.58}$. 
We rule out the possibility of it being a low-redshift interloper with the same argument supplied for system 11. 
A fourth image is expected at large radus but is not confidently identified due to its expected dimness.

System 13.
It comprises a pair of images straddling the critical curve, along with a very faint third image well beyond the tangential critical curves. 
The photometric redshift is z=1.39$\pm$0.23. 

System 14.
It was first identified by \citet{atek14} as their system 1, which comprises two images. 
We propose an alternative, which has a more plausible surface brightness and colour, to their second image. 
It is designated as 14.2 and is the most magnified image of system 14. 
We calculated for this image a photometric redshift of z=5.61$\pm$0.65. 
We also identified a third image, 14.3, which is badly buried in the light of a group of member galaxies but clearly identified at the expected location given by our model. 

System 15.
It consists of three images at a photometric redshift of 4.62$\pm$0.74, consistent with our geometric redshift of 4.86$\pm$0.21. 
Note that images 15.1 and 15.2 were identified by \citet{atek14} (their images 3.1 and 3.2), but the third image 15.3 has not previously been identified. 

System 16.
It was first identified by \citet{atek14} as having three images.
Our image 16.3 is firmly preferred by our lens model to the identification 4.1 proposed by \cite{atek14}, and with a colour consistent with 16.1 and 16.2. 
We derive a photometric redshift of z=4.89$\pm$0.58 from the brightest image 16.1.

System 17.
It is at very high redshift with a measured photometric redshift of z=6.65$\pm$0.85. 
Two central images lie close to the BCG. 
In this region a faint counter image claimed by \citet{atek14} (as their image 5.4) does not seem to correspond to a faint red source, but rather to noise. 
We do find the anticipated image relatively nearby, 17.4, and with consistent colours which we take to be a very secure identification. 

\begin{figure}
\centering
\includegraphics[width=85mm]{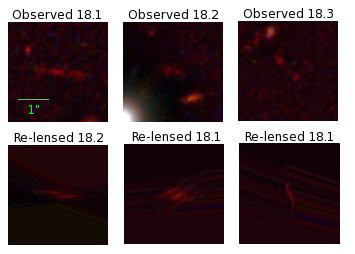}
\caption{
Delensing and relensing system 18 - a triply-lensed system with a photometric redshift of $\sim$7.01. 
The top row shows the observed images of system 18 and for comparison the bottom row shows selected relensed images. 
}
\label{system18}
\end{figure}

System 18. 
Triple image at photometric redshift of z=7.01$\pm$0.78, consistent with the finding of \citet{zheng14}. 
This high redshift system is unique in colour and detected only in the NIR bands. 
As shown in figure \ref{system18}, we find good agreement between the observed images and those we predict. 

System 19. 
This is a double image pair identified by \citet{zitrin14}, and is the multiply-lensed galaxy with the highest redshift in A2744. 
We adopt the photometric redshift of $z=9.83^{+0.22}_{-0.44}$ from \citet{zitrin14}. 
The unidentified third image may be too dim to be observed, or it is too heavily buried in the light of nearby galaxies. 

System 20.
It is identified in the complete HFF data as a triply-lensed blue galaxy with a long faint tail. 
Its photometric redshift is found to be z=3.38$\pm$0.43. 

System 21. 
Also, identified in the complete HFF data, it is a high-redshift triply-lensed galaxy at a photometric redshift of z=6.97$\pm$0.78. 
20.1 can be clearly seen in the Western side of the central BCG, while 20.2 appears close to one of the member galaxy, but is still visible. 
20.3 is predicted to appear at a large radius near the position of 17.1, but the expected flux is too low for it to be detected. 

System 22. 
It is green in colour, with two images straddling the critical curve near the central BCG, and a third image at large radius. 
Delensing and relensing this system yields the largest positional offset. 
We suspect this may be due to the relatively poor fit of the SED, which yields a photometric redshift of z-4.64$\pm$0.58. 

System 23. 
A blue galaxy with bimodal internal structure. 
A double pair can be clearly seen near images 1.1 and 1.2, but the 3rd image is predicted to have a small flux which is difficult for us to match. 
We derive a photometric redshift of z=1.81$\pm$0.28.

\subsection{Lens model}\label{sec:results}
 
\begin{figure*}
\centering
\includegraphics[width=150mm]{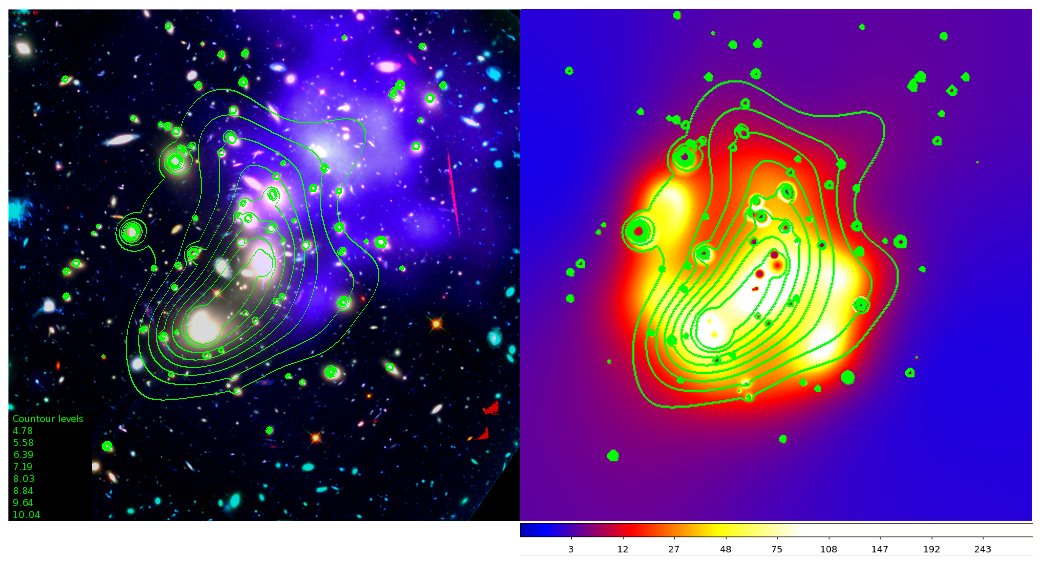}
\caption{
Left: Linearly spaced contours of our lens model overlaid on our colour image of A2744. 
The X-ray image obtained by Chandra is also overlaid as false violet colour. 
The offset gas emission indicates a major merger has recently taken place and the direction of this is consistent with the large nearby cluster just outside of the field of view towards the NW \citep{merten11}. 
The units of the surface mass density contours are $10^{8}$ $M_{\odot}$ $kpc^{-2}$, and the field of view is 2.56"$\times$2.56". 
Right: same set of mass contours overlaid on the mass S/N map. 
Most of the mass contours lie within the region where S/N is greater than 3. 
}
\label{mass}
\end{figure*}

\begin{figure}
\includegraphics[width=85mm]{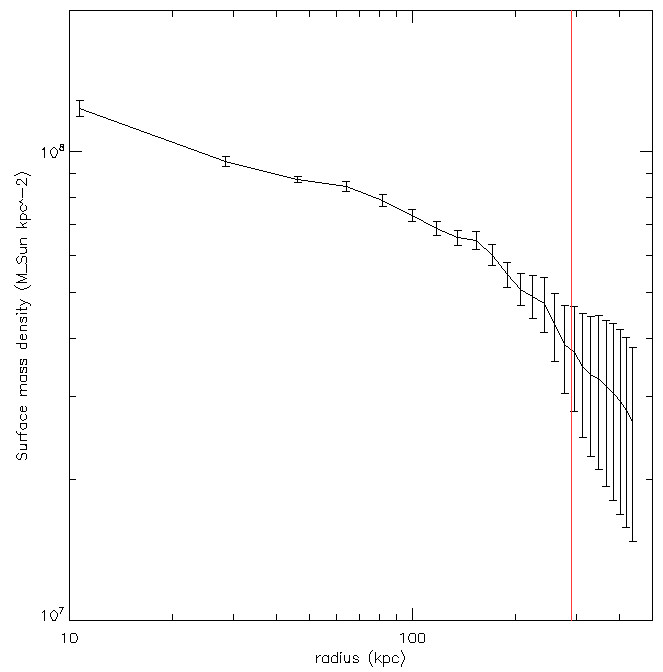}
\caption{
Radial profile of the surface mass density, plotted with circular annuli centred on the central BCG. 
The red vertical line marks the furthest multiply-lensed image. 
Error bars represent 1 standard deviation derived from a range of models. 
}
\label{radial_profile}
\end{figure}

\begin{figure}
\includegraphics[width=85mm]{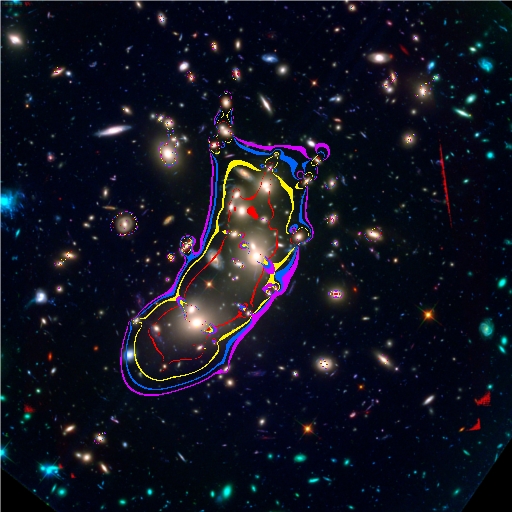}
\caption{
Critical curves derived from our lens model. 
They correspond to a range of redshift: z=1 (red), z=2 (yellow), z=4 (blue), and z=9 (magneta). 
The critical curves are selected as area in the magnification map with values greater than 200. 
}
\label{critical_curves}
\end{figure}

The solution and uncertainty we obtain for the mass distribution is shown in figure \ref{mass}, its radial profile in figure \ref{radial_profile}, and the corresponding critical curves are shown in \ref{critical_curves}. 
Apart from small perturbations associated with member galaxies, the overall mass distribution is smooth. 
There is a clear tendency for the smooth grid component to follow the member galaxy distribution in terms of the overall shape of the mass contours. 
The dark matter distribution is "boomerang" shaped with a southerly component seems to be elongated in the NE-SW direction coincident with the major axis of one of the two luminous cD galaxies, and the other component in the NW-SW direction approximately associated with the upper luminous galaxy. 
The elongation and bimodality of the cluster makes the radial profile not as revealing as that of relaxed clusters on the characteristics of mass distribution, but nevertheless provide a convenient comparison with other lens models. 
Interestingly, this upper luminous member galaxy seems to show evidence of an internal density wave structure affecting the observed stellar distribution. 
We plan to explore more thoroughly the dark matter distribution of this galaxy thanks to the presence of the two long radial arcs in system 2.

Interestingly, at the "apex" of the "boomerang", there is bright X-ray emission as can be seen in Figure \ref{mass}. 
This X-ray feature was reported in \cite{merten11} for which an interpretation was proposed involving multiple components and also highlighted in \citet{owers12} as a possible Bullet-like gas feature. 
We prefer to interpret it as simply one of several over-dense regions of gas that appear to comprise the generally very disturbed region of X-ray emission visible in Figure \ref{mass}. 
Plausibly the "excess" of mass we see at the apex of the boomerang is significantly contributed to by the X-ray emitting gas, in addition to dark matter. 
To explore possible scenarios quantitatively, a full hydrodynamical/N-body model is required, such as those used to model the collision of gas and dark mater in the bullet cluster \citep{mastropietro08, springel07, molnar13a} and other binary interacting systems such as A1750 \citep{molnar13b} and the very massive "El Gordo" colliding cluster \citep{molnar14}.

To demonstrate our accuracy we relens each observed image as a set of counter images for comparison with the observed images (as shown in figures \ref{stamps1} to \ref{stamps6}), and we mark the predicted centroids of these model images on the stamp for each counter image identified. 
There are between 1 to 3 centroids per observed image to compare with, depending on the number of counter images comprising each system. 
The generally high level of accuracy that we find in this comparison is clear, which in a lot of cases corresponds to only $\sim$0.2"-0.4" uncertainty in the relensed locations of counter images (figure \ref{histogram}). 
The RMS uncertainty is $\sim$1.25", which is slightly larger than that of a recent parametric model of A2744 ($\sim$0.69") that is constructed with double amount of constraints \citep{jauzac14}. 
We have also confidently corrected counter-images images that have been seemingly misidentified \citep{atek14} in systems 14 (14.2), 16 (16.3) and 17 (17.4). This brings to a total 65 images, corresponding to 21 multiply lensed systems that we are fully confident of.

We compare here the distances of the lensed systems inferred in the manner described in section \ref{sec:geoz} with their spectroscopic or photometric redshifts. 
The angle between the lensed image and the original source position scales linearly with the ratio of $d_{ls}/d_s$. 
This means that higher redshift systems are deflected by increasingly larger angles. 
A clear example of the distance dependence can be seen by simply comparing system 17 and system 2, which must lie very close to each other in the source plane, behind the BCG, as their multiple images are nearly coincident in the lens plane. 
It is clear here that the outer images of system 17 are deflected by significantly larger angles than for system 2 and that this difference is estimated accurately for in our model, arising from the wide difference in redshift which we estimate to be z=2.22$\pm$0.03 and z=6.04$\pm$0.77 geometrically and in good agreement with the independently determined photometric redshifts of 2.49$\pm$0.34 and 6.75$\pm$0.76.

\begin{figure}
\includegraphics[width=85mm]{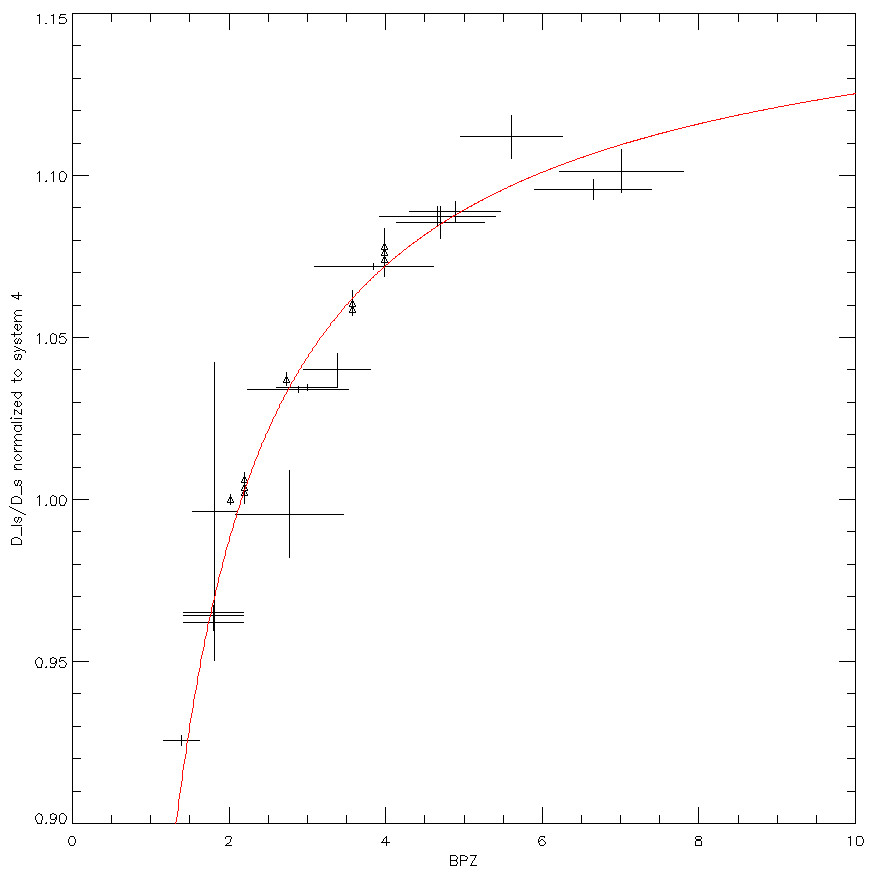}
\centering
\caption{
This figure shows the lensing distance ratio $f_{k}$ (D$_{ls}$/D$_{s}$, see equation 5) derived by applying our model to the multiple images we have detected. 
We have chosen to normalise the model by the distance ratio of system 4 as its spectroscopic redshift is secure. 
The remainder of the systems are plotted against photometric redshifts. 
We have excluded system 3 from this plot as the positional accuracy required is much too great to be useful for this purpose. 
21 systems were used to derive the lens model because they have relatively good SED fits and reliable photometric redshifts.  Systems 2, 3, 6 also have spectroscopic redshifts and are shown as triangle as with system 4. 
The curve corresponds to the lensing angular diameter distance-redshift relation derived for the current best values of $\Omega_\Lambda$ and $\Omega_m$.  
Error bars of the y-axis represent the standard deviation of six model reconstructions each with different initial conditions. 
Error bars of the x-axis represent 2-sigma dispersion in the probability distribution function in redshift space for the photometric redshifts.
}
\label{distance_bpz}
\end{figure}

\begin{figure}
\centering
\includegraphics[width=85mm]{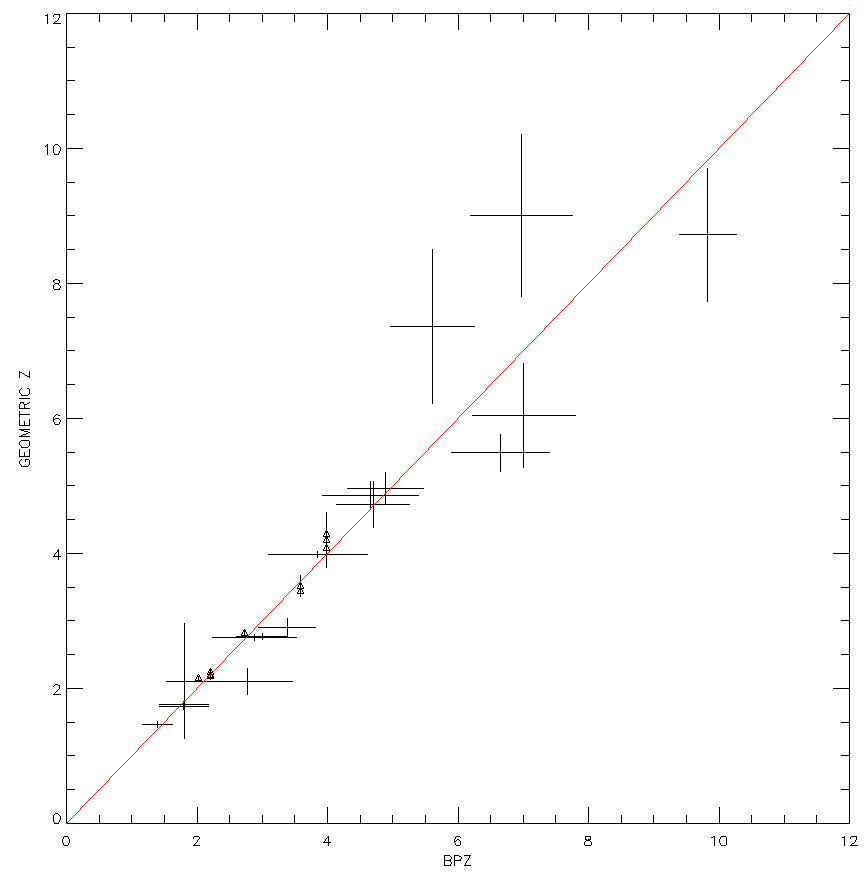}
\caption{
This is the same as the previous figure with the best fit lensing distances converted to redshifts, which we term "geometric redshifts" and we plot them against the photometric redshifts determined from the BPZ method.
Systems with spectroscopic redshifts are indicated by triangles. 
}
\label{geoz_bpz}
\end{figure}

To estimate the geometric redshift, we simply take our best fit lens model and derive the best distances by minimizing the separation of each set of delensed images in the source plane.
Figure \ref{distance_bpz} shows the derived ratio of angular diameter distances between the lens to the source and the observer to the source, normalized by the same ratio to system 4 (which has an accurate spectroscopic redshift), for all the lensed sources used to construct our lens model. 
Also plotted in this figure is the same ratio predicted theoretically for different choices of the cosmological parameters $\Omega_\Lambda$ and $\Omega_m$. 
There is a good agreement between the derived distance scaling factor and that predicted theoretically over a broad range of photometric redshifts for the lensed sources spanning z$\sim$1 to z$\sim$7. 
The trend seen in Figure \ref{distance_bpz} was first seen in the model of A1689 \citep{broadhurst05}, thus providing a useful check of the lens model. 
A significant scatter, however, was found and attributed to the simplicity of the lens model built on the assumption that mass approximately traces light. 
Here we also see a level of residual scatter, but significantly smaller than for A1689. 
This `best distances' plot can also be converted via a set of assumed cosmological parameters to a plot of geometric redshifts versus input redshifts.
Such a plot is shown in figure \ref{geoz_bpz}, there the relatively small scatter demonstrates the high level of self-consistency in our lens model. 
We also obtained realistic uncertainties for the lensing distances by starting the reconstruction with a combination of input parameters which vary by a factor of two smaller and larger about a sensible set of values,  including the initial mass-to-light ratios of the member galaxies and the initial mass contained in each cell in the grid component.

The level of self-consistency found in our free-form model is highly encouraging, given that the derived cluster mass distribution is smooth (figure \ref{mass}) with only small scale irregularities from cluster members, and hence clearly does not overfit the data. 
The precision achieved in the derived distance scaling parameter motivates an examination of the cosmological constraints possible, by solving simultaneously for both the mass distribution and cosmological parameters in a model-independent way. 
Mass distribution of the cluster A1689, $\Omega_m$ and $w_{x}$ have been attempted to be jointly constrained by a parametric model by \citet{jullo10}, which combined with WMAP and X-ray observation yield constraints competitive to other methods. 
Constraints on cosmological parameters with free-form lens models has been shown to be feasible with simulated lenses \citep{lubini14}. 
Application on real data, however, has yet to be realized.

\begin{figure}
\centering
\includegraphics[width=85mm]{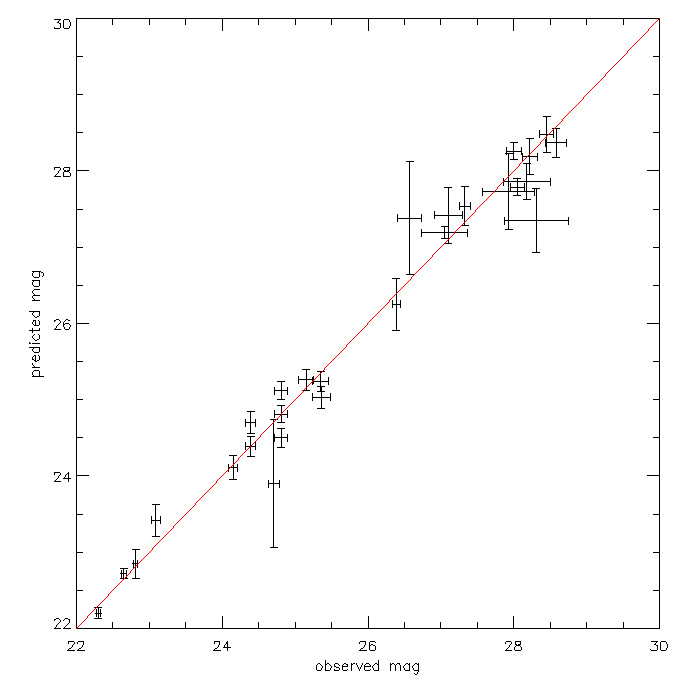}
\caption{
Relative magnitudes predicted by our model plotted against observed magnitudes of multiple images. 
The predicted magnitudes are calculated by magnifying the observed magnitudes by our lens model. 
The horizontal error bars represent the uncertainty in photometry while the vertical error bars are contributed by both photometric and model uncertainties. 
The rms uncertainty is about 0.25 magnitudes. 
The best fit line has a slope of 0.999$\pm$0.013 and a y-intercept of 0.024$\pm$0.276, which makes it indistinguishable from the perfect one-to-one red line. 
}
\label{mag}
\end{figure}

The ultimate goal of the lens models for the HFF program is achieve a reliable and precise correction for the magnification of the distant magnified galaxies so that intrinsic properties of lensed galaxies can be inferred. 
In figure \ref{mag} we plot the model-predicted relative magnitudes against the observed magnitudes for the images that constrained the reconstructed lens model. 
The magnitudes are predicted by magnifying or de-magnifying the observed magnitudes of the first and second images in each system, as 
typically there are three images per system with usually one case where the photometry is poor due to overlap with a member galaxy. 
Reassuringly, we find a clear linear relation between the predicted and observed magnitudes (slope=0.999$\pm$0.013; y-intercept=0.024$\pm$0.276), with some outliers which are attributable to the proximity to the critical curves but in general the scatter is small (rms$\sim$0.25 magnitude) and the trend shows no systematic deviation from linearity. 
Note that even though relative image brightness has not been used to constrain the lens model, the scatter in this plot is encouragingly small, which implies luminosity functions at high redshifts behind cluster lenses can now be accurately determined.

\section{Discussion}\label{sec:discussion}

The mass distribution we derive is smooth and generally follows the distribution of the bright member galaxies, but with some departures in detail. 
In figure \ref{mass} we can see a clear hint that the X-ray emission from three of the brighter emission regions seem to be affecting locally the mass map we derived from lensing. 
More generally, the offset between the centroid of our DM map the bulk of the X-ray emission is highly indicative of a major head-on collision along an axis connecting the DM distribution and the NW cluster component that lies $\sim$2' beyond the edge of our field, which is noticeable in the corner of the Hubble data by virtue of strongly lensed images present there - see also \cite{merten11}. 
The disturbed hot gas lies in between the two massive clusters along the axis joining them and closer to the main mass in our map, indicating this is the larger of the two mass components, similar to the case of the ``bullet cluster''. 
In light of our findings, a binary collision solution is motivated using the combined DM and hot gas information modelled with a combined hydrodynamical/N-body model, which has proved useful for several other major post-merger binary clusters including the ``bullet cluster'' and ``El Gordo'' \citep{mastropietro08, springel07, molnar14} and most recently for the HFF cluster MACS0416 \citep{diego14b}. 

We have found that the redshift distribution of multiply lensed images extends over a wide redshift range, from z$\sim$1.0 to z$\sim$10.0, and is fairly uniform in the high z range, so far as we may conclude on the basis of 21 sources. 
Beyond this nothing compelling is yet found despite the high magnification for multiply-lensed images generated by A2744. 
So it is interesting that the HFF filters set does permit the detection of more distant objects, to z$\sim$11.5, and yet despite the high magnification generated by this cluster there is only a few multiply-lensed galaxies above z$\sim$7. 
Conceivably, we may establish IR dropout galaxies more confidently with the addition of the upcoming optical data amongst the fainter NIR detections, but even with the full data set we do not anticipate significant additional numbers of multiply lensed galaxies with $z>7.5$ as very few very red NIR detections are unmatched within the critically lensed region.

This relative deficit of high-z galaxies is consistent with the conclusions for the single, less magnified high-z images identified photometrically by \citet{zheng14}, with careful photometry combined with HFF and Spitzer imaging. 
In this study, there also seems to be a similar effective upper redshift limit of z$\simeq$9, although the magnification of these single images is generally several times smaller than the highly magnified area where the multiply lensed sources lie. 
Clearly it is now important to model this redshift distribution and the corresponding luminosity functions with careful consideration of the selection effects so that the reality and significance of this deficit can be quantified. 
A recent calculation by \citet{coe14} estimated 6 objects are expected in a typical HFF field at z $>$ 9 based on the luminosity functions extrapolated from that observed in $4 < z < 8$ convolved with the magnification maps submitted by the community.
Clearly we will need to model the expected redshift distribution carefully to quantify the level of this deficit and we need to examine the five additional deep HFF clusters being scheduled, allowing for a more accurately defined redshift distribution averaged over large scale fluctuations in galaxy numbers.

Monotonously declining evolution is of course expected in the gravitationally scale free case of standard collisionless particle interpretation of CDM, and this is probably consistent with the measured evolution of the integrated star formation rate of \citet{oesch13}, at $z<8$, given the freedom to relate the observables to the predicted growth of the halo mass function. However, a sharp break above $z=8$ does look like a surprising departure in this context and may hint that the process of galaxy formation is not scale free as with CDM, but begins more suddenly than expected. 
Or perhaps, the reionization period occurred, on average, at later times than suggested by recent polarization data of CMB, and more in line with measurements of the Lyman$\alpha$ forest. 
The sense of this difference may point towards a lack of small halos at this simplest level, and is certainly not in conflict with local observations of the galaxy mass function which does not seem to extend much below a few $\times 10^7M_\odot$, as judged by the local dwarf galaxies \citep{strigari12}. 
Warm dark Matter (WDM) may help reconcile this behaviour, where free streaming of previously relativistic dark matter is invoked to suppress small structures and provide shallow cores. 
However, detailed calculations of this do not seem to manage to be self-consistent, as the light particle mass of $\sim$0.5 keV required to generate large cores, would eliminate the formation of too many galaxies, for which a higher minimum mass of $\sim$3 keV is required to permit the formation of dwarf spheroidal galaxies \citep{schneider13}. 

Another possibility has emerged from consideration of cold dark matter in condensate form where the first simulations of this
unexplored form of CDM show how the Jeans scale inherent to a condensate can sharply suppress objects below $10^8M_\odot$, and provide ``solitonic'' cores of constant density in dwarf spheroidal galaxies, without preventing the formation of more massive galaxies. 
The precise redshift evolution predicted by this wavelike form of dark matter, termed $\psi$DM by Schive, Chieuh $\&$ Broadhurst (2014) will be very interesting to compare with the emerging HFF data at high redshift with potentially profound implications for the nature of dark matter.

\section{Summary, Conclusions and Future Work}\label{sec:conclusions}

By utilising cluster magnification the HFF program provides unprecedentedly deep optical-NIR imaging for detecting galaxies to higher redshifts and lower luminosities. 
The clusters chosen for this program are those established to have large sky areas of high magnification (\citet{kneib93, zitrin09b, merten11, zitrin13} for the clusters A370, MACS1149, A2744, MACS0416 respectively).  
To derive the physical properties of the distant galaxies detected it is of course essential that an accurate lens model can be constructed to reliably correct for the lens magnification over the field of view. 
In this paper we have demonstrated that our "free-form" lensing method, WSLAP+, has significantly improved the ability to securely identify multiply lensed images in deep imaging data, allowing us to uncover 4 new systems with confidence and to identify errors in previous work based on the same HFF data. 
Typically even with the best Hubble data a large proportion of faint arcs remain unmatched to counter-images, even with considerable modelling efforts. 
A significant part of this problem is the relative inflexibility of parameterised models. 
Such models can only be partially appropriate at best for cluster mass distributions, particularly in the case of merging clusters where the complexities of tidal effects during encounters means the general mass distribution cannot be expected to adhere to a sum of idealised elliptical, power-law mass halos usually adopted. 

This positional insecurity, and the increased quality of the new deep HFF data has motivated our renewed examination of free-form modelling, augmented by the deflections from member galaxies. 
We simply combine a Gaussian pixel grid for describing a smooth cluster component, together with small scale perturbations from the observed member galaxies. 
It is the combination of these high and low frequency components which we find provides meaningful solutions when searching for multiple images, as typically one or more counter-images in any set of multiply lensed images is locally perturbed or created by the presence of a member galaxy. 
We have tested our method previously with realistic simulations \citep{sendra14} and we have applied it to the rich dataset of A1689 where we found 12 new multiple systems of images and improved the resolution of the recovered mass distribution \citep{diego14}.

This good agreement between photometric redshifts and geometric redshifts indicates a high degree of self-consistency, which other lens models of the same cluster have yet to demonstrate rigorously. 
The linearity enables us to estimate the redshifts of multiply lensed galaxies which were not used as constraints. 
This is illustrated in \citet{zitrin14}, in which our model excludes the possibility of the z=10 system being a low-redshift interloper. 
The free-form model-independence of our method can permit a joint constraint on the mass distribution and cosmology. 
The feasibility of this has been examined with simulations by \citet{lubini14}, and now seems warranted by our improved observational precision achieved here. 
We have also examined the self-consistency of the model by comparing the  predicted and observed brightnesses of the lensed images as shown in Figure \ref{mag}.

We conclude from the precision and self-consistency of our lens model that we have constructed a reliable Free-Form lensing model of A2744. 
It is apparent that this mass distribution is smooth in 2D, punctuated only by member galaxies, with no evidence of local perturbations that would otherwise imply over-constrained modelling. 
This model has allowed us to improve the reliability of multiply-lensed images with convincing identifications for 21 sets of multiply-lensed images, using the new deep NIR imaging from the HFF, combined with the existing optical/ACS data from \cite{merten11}.  
Our sample adds 4 new systems unknown prior to the HFF program and, furthermore, we correct multiple errors in previous work based on the HFF data. 
Further improvements will come with the upcoming deeper optical data for reducing the ambiguity in identifying counter images amongst the many faint blue images lying at $z<3.5$ for a complete derivation of the redshift distribution. 
It will also be very interesting to examine the constraints we can impose on the mass profile of the BCG galaxy for which the pair of fortuitously long radial arcs of system 2 provide perhaps the best current constraints the mass profile of this interesting class of galaxy and may help in a better understanding of the origin of cD galaxies in relation to more normal cluster members.

\acknowledgements
$\it{Acknowledgements.}$
The authors wish to thank the referee, Adi Zitrin, for very useful comments and suggestions that greatly helped improve this paper. 
The work is based on observations made with the NASA/ESA {\it Hubble Space Telescope} and operated by the Association of Universities for Research in Astronomy, Inc. under NASA contract NAS 5-26555. 
T. J. B. thanks the University of Hong Kong for generous hospitality. 
J. M. D. acknowledges support of the consolider project CAD2010-00064 and AYA2012-39475-C02-01 funded by the Ministerio de Economia y Competitividad. 

\clearpage

\clearpage

\LongTables
\begin{deluxetable}{cccccp{6cm}}
\centering
\tabletypesize{\footnotesize}
\tablecaption{Detailed information of individual lensed images}
\tablewidth{0pt}
\tablehead{\colhead{Image} & \colhead{RA} & \colhead{Dec} & \colhead{BPZ (image used)} & \colhead{Geo-z} & \colhead{Remarks}}
\startdata
1.1.1 & 3.5975951 & -30.403926 & 1.80$\pm$0.26 (1.3) & 1.75$\pm$0.02 &  \\
1.2.1 & 3.5959487 & -30.406827 & & & \\
1.3.1 & 3.5862093 & -30.410008 &  &  &  \\
1.1.2 & 3.5970415 & -30.404746 &  &  &  \\
1.2.2 & 3.5963711 & -30.406161 &  &  &  \\
1.3.2 & 3.5857255 & -30.410085 &  &  &  \\
1.1.3 & 3.5975204 & -30.403172 &  &  &  \\
1.2.3 & 3.5952961 & -30.406992 &  &  &  \\
1.3.3 & 3.5858301 & -30.409656 &  &  &  \\
1.1.4 & 3.5980799 & -30.40399 &  &  &  \\
1.2.4 & 3.5957233 & -30.407548 &  &  &  \\
1.3.4 & 3.5873867 & -30.410161 &  &  &  \\

2.1.1 & 3.5832661 & -30.403339 & $1.86^{+0.73}_{-0.28}$ (2.4) & 2.22$\pm$0.04 & $z_{spec}$=2.2 used as input.  \\
2.2.1 & 3.5864069 & -30.402130 &  &  & \\
2.3.1 & 3.5853776 & -30.399889 &  &  & \\
2.4.1 & 3.5972758 & -30.396723 &  &  & \\
2.1.2 & 3.5825262 & -30.402290 &  &  &  \\
2.2.2 & 3.5862188 & -30.400850 &  &  & \\
2.3.2 & 3.5844674 & -30.399292 &  &  & \\
2.4.2 & 3.5967284 & -30.396298 &  &  & \\
2.1.3 & 3.5830255 & -30.403189&  &  &  \\
2.2.3 & 3.586436 & -30.401876 &  &  & \\
2.3.3 & 3.5851302 & -30.39967 &  &  & \\
2.4.3 & 3.5971321 & -30.396639 &  &  & \\

3.1.1 & 3.5894856 & -30.39387 & 4.11$\pm$0.50 (3.1+3.2) & 4.19$\pm$0.27 & $z_{spec}$=3.98 used as input \\
3.2.1 & 3.5893682 & -30.39386 &  &  &  \\
3.3.1 & 3.5774772 & -30.39956 &  &  &  \\
3.1.2 & 3.589202 & -30.393845 &  &  &  \\
3.2.2 & 3.5887972 & -30.393803 &  &  &  \\
3.3.2 & 3.5775032 & -30.399459 &  &  &  \\
3.1.3 & 3.5892184 & -30.393849 &  &  &  \\
3.2.3 & 3.5889638 & -30.393823 &  &  &  \\
3.3.3 & 3.5775394 & -30.399379&  &  &  \\

4.1.1 & 3.5804456 & -30.408951 & 3.47$\pm$0.44 (4.3) & 3.49$\pm$0.11 & $z_{spec}$=3.580 is used as input \\
4.2.1 & 3.5921262 & -30.402667 &  &  &  \\
4.3.1 & 3.5956659 & -30.401634 &  &  &  \\
4.4.1 & 3.5937634 & -30.405167 &  &  &  \\
4.5.1 & 3.5931248 & -30.404871 &  &  &  4.4 $\&$ 4.5 identified additional to M11 \\
4.1.2 & 3.5802506 & -30.408751&  &  &  \\
4.2.2 & 3.5920986 & -30.402534 &  &  &  \\
4.3.2 & 3.5955637 & -30.401517 &  &  &  \\
4.4.2 & 3.5934919 & -30.40506 &  &  &  \\
4.5.2 & 3.5932915 & -30.404965 &  &  &  \\

5.1 & 3.5849754 & -30.391399 & $3.90^{+0.76}_{-0.59}$ (5.2) & 3.98$\pm$0.04 & \\
5.2 & 3.5834511 & -30.392062 &  &  &  \\
5.3 & 3.5801170 & -30.394659 &  &  &  \\

6.1 & 3.5864121 & -30.409342 & 2.34$\pm$0.35 (6.3) & 2.16$\pm$0.02 & $z_{spec}$=2.019 is used as input \\
6.2 & 3.5940639 & -30.407996 &  &  &  \\
6.3 & 3.5985646 & -30.401822 &  &  &  \\

7.1 & 3.5846077 & -30.40982 & $3.00^{+0.39}_{-2.80}$ (7.3) & 2.77$\pm$0.01 & \\
7.2 & 3.5952251 & -30.407406 &  &  &  \\
7.3 & 3.5982661 & -30.402331 &  &  &  \\

9.1 & 3.5883741 & -30.405267 & $1.74^{+1.34}_{-1.62}$ (9.2) &  & Not used in reconstruction \\
9.2 & 3.5871319 & -30.406217 &  &  & \\
(9.3) & 3.6014708 & -30.396005 &  &  &  alternative to M11's 9.3 \\

10.1 & 3.5884046 & -30.405882 & $2.85^{+0.39}_{-2.70}$ (10.3) & 2.76$\pm$0.02 &  \\
10.2 & 3.5873821 & -30.406479 &  &  & \\
10.3 & 3.6007145 & -30.397103 &  &  & \\

11.1 & 3.5914271 & -30.403917  & $2.88^{+0.64}_{-2.67}$ (11.2) & 2.76$\pm$0.02 & \\
11.2 & 3.5972432 & -30.401431 &  &  &  \\
11.3 & 3.5827069 & -30.408931 &  &  & \\
11.4 & 3.5945371 & -30.40655 &  &  & \\

12.1 & 3.5936217 & -30.404463 & $2.77^{+0.69}_{-2.58}$ (12.3) & 2.10$\pm$0.19 &  \\
12.2 & 3.5932342 & -30.403254 &  &  &  \\
12.3 & 3.5945698 & -30.402983 &  &  &  \\

13.1 & 3.5937765 & -30.402181 & 1.39$\pm$0.23 (13.1) & 1.47$\pm$0.01\\
13.2 & 3.5923693 & -30.402551 &  &  & \\
13.3 & 3.582771 & -30.408042 &  &  &  \\

14.1 & 3.5761424 & -30.404493 & 5.61$\pm$0.65 (14.2) & 7.36$\pm$1.14 & Corresponds to A14's system 1 \\
14.2 & 3.5907599 & -30.395578 &  &  & alternative to A14's 1.2 \\
14.3 & 3.5883651 & -30.395642 &  &  & Identified in addition to A14 \\

15.1 & 3.5935448 & -30.409717 & $4.62^{+0.74}_{-4.27}$ (15.1) & 4.86$\pm$0.21 & Corresponds to A14's 3.1 \\
15.2 & 3.6005113 & -30.40183 &  &  & Corresponds to A14's 3.2 \\
15.3 & 3.5881426 & -30.410567 &   &  & Identified in addition to A14 \\

16.1 & 3.5965504 & -30.409002 & 4.31$\pm$1.16 (16.1) & 4.96$\pm$0.24 & Corresponds to A14's 4.3 \\
16.2 & 3.6000445 & -30.404408 &  &  & Corresponds to A14's 4.2 \\
16.3 & 3.5879389 & -30.411597 &  &  & alternative to A14's 4.1 \\

17.1 & 3.5978104 & -30.395982 & 6.65$\pm$0.75 (17.2) & 5.49$\pm$0.28 & Corresponds to A14's 5.3 \\
17.2 & 3.5804218 & -30.405072 &  &  & Corresponds to A14's 5.1 \\
17.3 & 3.5853276 & -30.397933 &  &  &  Corresponds to A14's 5.2 \\
17.4 & 3.5874584 & -30.401369 &  &  & alternative to A14's 5.4 \\

18.1 & 3.5923072 & -30.409931 & 7.07$^{+0.78}_{-6.16}$  (18.1) & 6.04$\pm$0.77 &  \\
18.2 & 3.5884178 & -30.410324 &  &  & \\
18.3 & 3.6007742 & -30.400966 &  &  & \\

19.1 & 3.5925126 & -30.401484 & $9.83^{+0.22}_{-0.44}$   (19.1) &  &  \\
19.2 & 3.5950299 & -30.400752 &  &  & \\

20.1 & 3.5938804 & -30.409723 & 3.38$\pm$0.43  (20.1) & 8.72$\pm$0.98 &  \\
20.2 & 3.5903491 & -30.410578 &  &  & \\
20.3 & 3.6001002 & -30.402951 &  &  & \\

21.1 & 3.5798457 & -30.401595 & 6.97$\pm$0.78   (21.1) & 9.01$\pm$1.20 &  \\
21.2 & 3.5835462 & -30.396701 &  &  & \\

22.1 & 3.5859365 & -30.403161 & 4.70$\pm$0.56  (22.1) & 4.73$\pm$0.34 &  \\
22.2 & 3.5837153 & -30.404103 &  &  & \\
22.3 & 3.6006567 & -30.395436 &  &  & \\

23.1 & 3.5962145 & -30.403044 & $1.81^{+0.28}_{-1.65}$   (23.1) & 2.11$\pm$0.85 &  \\
23.2 & 3.5952243 & -30.405394 &  &  & \\

\enddata
\label{table1}
\end{deluxetable}

\begin{deluxetable}{ccccc}
\tabletypesize{\scriptsize}
\tablewidth{0pt}
\tablecaption{Magnification and photometry of individual lensed images}
\tablehead{\colhead{Image} & \colhead{$\mu$} & \colhead{Observed AB mag} & \colhead{Predicted magnitude(s)}}
\startdata
1.1 & 6.29$\pm$0.03 & 22.30$\pm$0.03 (F160W) & 22.20$\pm$0.07 \\
1.2 & 7.82$\pm$0.06 &  & \\
1.3 & 4.17$\pm$0.14 & 22.65$\pm$0.03 (F160W) & 22.75$\pm$0.07\\

2.1	 & 22.17$\pm$1.25 & 22.81$\pm$0.03 (F160W) & 22.85$\pm$0.19 \\
2.2  & 15.82$\pm$1.25 &  & \\
2.3	 & 8.56$\pm$1.07 &  & \\
2.4  & 6.72$\pm$0.71 &  24.15$\pm$0.06 (F160W) & 24.11$\pm$0.16 \\

3.1 & 32.85$\pm$1.96 & 23.09$\pm$0.06 (F160W, 3.1+3.2) & 23.42$\pm$0.21 \\
3.2 & 14.43$\pm$1.31 &  & \\
3.3 & 7.90$\pm$0.31 & 25.36$\pm$0.12 (F160W) & 25.03$\pm$0.15 \\

4.1 & 5.37$\pm$1.20 &  & \\
4.2 & 7.29$\pm$0.24 & &  \\
4.3 & 7.54$\pm$0.23 & & \\
4.4 & 5.82$\pm$0.25 &  & \\
4.5 & 3.03$\pm$0.12 &  & \\

5.1 & 50.33$\pm$13.63 & & \\
5.2 & 411.56$\pm$85.31 &  & \\
5.3 & 10.78$\pm$1.78 & & \\

6.1 & 5.19$\pm$0.19 & 24.39$\pm$0.07 (F160W) & 24.70$\pm$0.14, 24.39$\pm$0.13 \\
6.2 & 4.68$\pm$0.10 & 24.81$\pm$0.09 (F160W) & 24.50$\pm$0.12, 24.50$\pm$0.12 \\
6.3 & 3.51$\pm$0.05 & 24.81$\pm$0.09 (F160W) & 24.81$\pm$0.11, 25.12$\pm$0.12 \\

7.1 & 3.53$\pm$0.10 & 25.35$\pm$0.10 (F160W) & 25.24$\pm$0.13 \\
7.2 & 5.54$\pm$0.05 &  &  \\
7.3 & 3.82$\pm$0.03 & 25.15$\pm$0.10 (F160W) & 25.26$\pm$0.14 \\

10.1 & 30.39$\pm$2.12 &  & \\
10.2 & 64.08$\pm$83.6 & & \\
(10.3) & 4.88$\pm$0.19 & & \\

11.1 & 2.70$\pm$0.13 &  & \\
11.2 & 4.33$\pm$0.10 &  & \\
11.3 & 4.26$\pm$0.02 &  & \\

12.1 & 22.70$\pm$1.43 & & \\
12.2 & 25.86$\pm$0.80 &  & \\
12.3 & 46.30$\pm$2.21 &  & \\

13.1 & 17.26$\pm$0.81 &  &  \\
13.2 & 16.35$\pm$0.59 & & \\
13.3 & 3.92$\pm$0.06 & & \\

14.1 & 3.96$\pm$0.57 & 26.57$\pm$0.17 (F105W) & 27.38$\pm$0.74 \\
14.2 & 46.25$\pm$27.59 & 24.71$\pm$0.07 (F105W) & 23.90$\pm$0.84 \\
14.3 & 4.54$\pm$1.48 & & \\

15.1 & 7.82$\pm$0.20 & 27.10$\pm$0.19 (F814W) & 27.42$\pm$0.37 \\
15.2 & 3.88$\pm$0.16 & 28.18$\pm$0.32 (F814W) & 27.86$\pm$0.24 & \\
15.3 & 7.77$\pm$0.68 &  & \\

16.1 & 6.30$\pm$0.08 & 28.05$\pm$0.10 (F105W) & 27.79$\pm$0.11 \\
16.2 & 5.19$\pm$0.01 & 28.00$\pm$0.10 (F105W) & 28.26$\pm$0.11 \\
16.3 & 6.56$\pm$0.42 &  & \\

17.1 & 7.81$\pm$0.66 & 28.45$\pm$0.10 (F160W) & 28.48$\pm$0.24 \\
17.2 & 9.88$\pm$0.91 & 28.22$\pm$0.10 (F160W) & 28.19$\pm$0.24 \\
17.3 & 5.63$\pm$0.34 &  & \\
17.4 & 10.10$\pm$1.29 &  & \\

18.1 & 10.35$\pm$0.36 & 27.33$\pm$0.08 (F125W) & 27.54$\pm$0.26 \\
18.2 & 11.69$\pm$0.86 &  & \\
18.3 & 3.97$\pm$0.36 & 28.58$\pm$0.15 (F125W) & 28.37$\pm$0.19 \\

19.1 & 17.79$\pm$0.92 & 27.93$\pm$0.36 (F160W) & 27.73$\pm$0.50 \\
19.2 & 10.41$\pm$0.13 &  28.31$\pm$0.44 (F160W) & 27.35$\pm$0.42 \\

20.1 & 9.57$\pm$0.15 & 26.39$\pm$0.05 (F606W) & 26.25$\pm$0.34 \\
20.2 & 12.73$\pm$0.95 &  & \\
20.3 & 4.57$\pm$0.11 & 27.05$\pm$0.31 (F606W) & 27.19$\pm$0.08 \\

21.1 & 15.95$\pm$1.37 &  &  \\
21.2 & 11.03$\pm$0.41 & & \\

22.1 & 51.34$\pm$6.75 &  & \\
22.2 & 41.89$\pm$4.51 &  & \\
22.3 & 2.59$\pm$0.23 &  & \\

23.1 & 7.84$\pm$0.06 &  &  \\
23.2 & 8.94$\pm$0.09 &  & \\

\enddata
\tablecomments{Only those systems with more than two photometry data have their observed and predicted magnitudes listed. }
\label{table2}
\end{deluxetable}


\end{document}